\documentclass[aps,prd,longbibliography,preprintnumbers,nofootinbib,
superscriptaddress,amsmath,floatfix]{revtex4}

\usepackage{amssymb}  
\usepackage{graphicx,subfigure} 
\usepackage{exscale}
\usepackage{textcomp}
\usepackage{enumerate}
\usepackage [english]{babel}
\usepackage[utf8]{inputenc}
\usepackage [autostyle, english = american]{csquotes}
\usepackage{slashed}
\usepackage{hyperref}
\usepackage{color}
\usepackage{float}
\definecolor{darkgreen}{rgb}{0.0,0.5,0}

\usepackage{color}

\usepackage[normalem]{ulem}  

\DeclareMathOperator{\Tr}{Tr}

\begin{document}

\title{SU(3) parity doubling in cold neutron star matter}

\author{Eduardo S.\ Fraga}
\email{fraga@if.ufrj.br}
\affiliation{Instituto de F\'{\i}sica, Universidade Federal do Rio de Janeiro, 
Caixa Postal 68528, 21941-972, Rio de Janeiro, RJ, Brazil}

\author{Rodrigo da Mata}
\email{rsilva@pos.if.ufrj.br}
\affiliation{Instituto de F\'{\i}sica, Universidade Federal do Rio de Janeiro, 
Caixa Postal 68528, 21941-972, Rio de Janeiro, RJ, Brazil}
\affiliation{Institut für Theoretische Physik, Goethe Universität, Max-von-Laue-Straße 1, 60438 Frankfurt am Main, Germany}

\author{Jürgen Schaffner-Bielich}
\email{schaffner@astro.uni-frankfurt.de}
\affiliation{Institut für Theoretische Physik, Goethe Universität, Max-von-Laue-Straße 1, 60438 Frankfurt am Main, Germany}



\begin{abstract} 

We present a phenomenological model to investigate the chiral phase transition characterized by parity doubling in dense, beta equilibrated, cold matter. Our model incorporates effective interactions constrained by SU(3) relations and considers baryonic degrees of freedom. By constraining the model with astrophysical data and nuclear matter properties, we find a first-order phase transition within realistic values of the slope parameter L. The inclusion of the baryon octet and negative parity partners, along with a chiral-invariant mass $m_{0}$, allows for a chiral symmetric phase with massive hadrons. Through exploration of parameter space, we identify parameter sets satisfying mass and radius constraints without requiring a partonic phase. The appearance of the parity partner of the nucleon, the N(1535) resonance, suppresses strangeness, pushing hyperonization to higher densities. We observe a mild first-order phase transition to the chirally restored phase, governed by $m_{0}$. Our calculations of surface tension highlight its strong dependence on $m_{0}$. The existence of mixed phases is ruled out since they become energetically too costly. We compare stars with metastable and stable cores using both branches of the equation of state. Despite limited lifespans due to low surface tension values, phase conversion and star contraction could impact neutron stars with masses around 1.3 solar masses or more. 
We discuss some applications of this model in its nonzero temperatures generalization and scenarios beyond beta equilibrium that can provide insights into core-collapse supernovae, protoneutron star evolution, and neutron star mergers. Core-collapse supernovae dynamics, influenced by chiral symmetry restoration and exotic hadronic states, affect explosion mechanisms and nucleosynthesis.

\end{abstract}

\maketitle


\section{Introduction}

First principle calculations on the lattice \cite{Fodor:2001pe,Aoki:2009sc} support the idea that hadronic matter undergoes a transition to a chirally restored phase known as the quark-gluon plasma at high temperatures and low baryon densities. Recently, the existence of yet another type of chirally symmetric phase dominated by chromoelectric interactions has been suggested \cite{Glozman:2022lda}. Despite these expectations, the precise nature and location of this transition in the phase diagram remain uncertain. When dealing with finite chemical potentials, Monte Carlo techniques on the lattice are hindered by the notorious sign problem \cite{deForcrand:2009zkb,Gattringer:2016kco}. Consequently, adopting effective models becomes the practical approach to capture the microphysics of the relevant degrees of freedom. Perturbative QCD at finite densities is also an option but can only be safely applied for extremely large densities, surpassing the typical central densities of compact stars \cite{Kurkela:2014vha,Fraga:2013qra,Fraga:2015xha,Gorda:2023mkk}, which are the densest known physical systems in the universe.

In this study, our focus is directed to the chiral phase transition occurring at high baryochemical potentials and zero temperature. We specifically examine how the interplay between parity doubling and hyperonization influences the properties of neutron star matter. Chiral approaches are employed, introducing a nonvanishing nucleon mass that remains finite even during the restoration of chiral symmetry. This is achieved through the mirror assignment of chirality within the parity doublet model \cite{lee1972chiral}.

Two-flavor parity doubling models have found extensive applications, ranging from investigations of vacuum phenomenology \cite{Nemoto:1998um,PhysRevD.82.014004,PhysRevD.85.054022} to in-medium studies of hot and dense baryonic matter \cite{HATSUDA198911,PhysRevC.75.055202,PhysRevC.77.025803,PhysRevC.82.035204,PhysRevC.84.045208,PhysRevC.87.015804,PhysRevD.92.054022,PhysRevC.92.015214,PhysRevD.91.125034,sym15030745,Koch:2023oez}, with particular relevance to neutron stars physics. Despite the depth of exploration in the context of two flavors, investigations of three-flavor scenarios remains relatively scarce.

Here, we address the effects of hyperonization and parity doubling within neutron star composition. Our approach is built upon a prior model \cite{Fraga:2022yls}, where higher orders of the scalar potential proved to be necessary for a good description of nuclear physics and also to avoid the Lee-Wick instability \cite{PhysRevD.9.2291} without a dynamically generated vector meson mass \cite{BOGUTA198334}. Our goal is to incorporate a chirally invariant mass enabled by the inclusion of a negative parity baryon octet. This understanding is pivotal for comprehending the properties of compact stars that may potentially contain chirally restored matter. To study dense nuclear matter and investigate the chiral phase transition characterized by parity doubling, we have developed a phenomenological model that incorporates baryonic degrees of freedom and effective interactions, with constraints derived from SU(3) relations. By constraining the model with astrophysical data and properties of nuclear matter, we find that 
a first-order phase transition for reasonable values of the slope parameter L can occur in the core of neutron stars.

The incorporation of parity doubling in our model allows for stable static configurations of stars with a metastable matter core, enabling stars with masses higher than the expected minimum mass of a neutron star formed via core-collapse supernova \cite{Suwa:2018uni} and around the value of the less massive observed neutron star \cite{Tauris:2019sho} which makes metastability related phenomena particularly relevant. A key advantage of using a unified model that describes both the neutron matter phase with fully broken chiral symmetry and the approximately chirally restored phase is the ability to calculate the surface tension and to determine, without external input, the location of the phase transition and also assess the possibility of mixed phases. This information is essential for determining the timescales of phase conversion via nucleation and understanding the potential signals emitted by compact objects that can, potentially, undergo such a phase transition.

This work is organized as follows: in Sec. \ref{sec:setup}, we outline the formalism. In particular, we briefly review the SU(2) parity doublet model in Sec \ref{subsec:su(2)}, which serves as a warm up for the SU(3) model discussed in  Sec. \ref{subsec:su(3)}. The complete Lagrangian is discussed in Sec. \ref{completeL}, stationary equations in Sec. \ref{subsec:freeenergy}, and a discussion on how to fix the parameters is left to Sec. \ref{subsec:parameterfixing}. We present our results in Sec. \ref{sec:results}. The final section, Sec. \ref{sec:summary}, summarizes our findings and points to possible future investigations and improvements within this approach.


\section{The model}
\label{sec:setup}
\subsection{Parity doublets in SU(2)}\label{subsec:su(2)}
 
Our approach is based on the ``mirror fermion" field formalism \cite{lee1972chiral,montvay1987chiral,Nemoto:1998um}, in which negative and positive parity fermion chiral transformation properties are reversed. As an example, take the nucleon fields under $SU(2)_{L} \times SU(2)_{R}$
\begin{equation}
    \psi_{+ L} \rightarrow \psi^{'}_{+ L} = L\, \psi_{+ L}, \hspace{1cm} \psi_{+ R} \rightarrow \psi^{'}_{+ R} = R \psi_{+ R} \, .
\end{equation}
Then, we assume
\begin{equation}
    \psi_{- L} \rightarrow \psi^{'}_{- L} = R\, \psi_{- L}, \hspace{1cm} \psi_{- R} \rightarrow \psi^{'}_{- R} = L \psi_{- R} \, .
\end{equation}
For these fields, one can construct the following chirally invariant mass terms
\begin{equation}\label{lagrangiansu2}
\mathcal{L}_{mass}^{inv}= -i m_{0}\bar{\Psi} \rho_{2} \gamma_{5} \Psi -g_{1} \bar{\Psi} ( \sigma +i \vec{\pi} \cdot \vec{\tau} \rho_{3}\gamma_{5})\Psi -g_{2} \bar{\Psi} (\rho_{3} \sigma +i \vec{\pi} \cdot \vec{\tau} \gamma_{5})\Psi \, ,
\end{equation}
where $\Psi$ is a doublet in parity space over which the Pauli matrices $\rho_{i}$ act upon. Here, we also have the sigma field and pions. The term proportional to $m_0$ mixes parity fields, giving rise to a nonzero nucleon mass even when chiral symmetry is restored.

At mean field level and defining $ \sigma \equiv \langle \sigma \rangle$, the part of the Lagrangian corresponding to the mass of the fields reads
\begin{equation}
\mathcal{L}_{mass}^{inv} =\left(\begin{array}{cc} \bar{\psi}_{+}, & \bar{\psi}_{-}
\end{array}\right) \cdot \left[\begin{array}{cc} (g_{1}-g_{2})\sigma & m_{0} \gamma_{5} \\ -m_{0} \gamma_{5} &
(g_{1}+g_{2})\sigma
\end{array}\right] \cdot \left(\begin{array}{cc} \psi_{+}\\  \psi_{-}
\end{array}\right) \, .
\end{equation}
It is convenient to rewrite the Lagrangian in terms of mass and parity eigenstates. This can be accomplished by the transformation that diagonalizes the fields in parity space defined by:
\begin{align}\nonumber
\Psi^{'} &= \frac{1}{\sqrt{2 \cosh{\delta}}} \left(\begin{array}{cc} e^{\delta/2} & e^{-\delta/2} \gamma_{5} \\
e^{-\delta/2} \gamma_{5} & e^{\delta/2}
\end{array}\right) \Psi \, , \\ 
\bar{\Psi}^{'} &= \bar{\Psi} \frac{1}{\sqrt{2 \cosh{\delta}}} \left(\begin{array}{cc} e^{\delta/2} & -e^{-\delta/2} \gamma_{5} \\
-e^{-\delta/2} \gamma_{5} & e^{\delta/2}
\end{array}\right) \, ,
\end{align}
with $\delta$ implicitly defined by $\sinh{\delta}=g_{1}\sigma/m_{0}$. In terms of the new defined fields, we have
\begin{equation}
\mathcal{L}_{mass} =\bar{\Psi}^{'} \cdot \left[\begin{array}{cc} m_{+} & 0 \\0 &
m_{-}
\end{array}\right] \cdot \Psi^{'} \, ,
\end{equation}
where
\begin{equation}
    m_{\pm}= \pm g_{2} \sigma + \sqrt{(g_{1}\sigma)^{2}+m_{0}^{2}}\, . 
\end{equation}
%

\subsection{Parity doublets in SU(3)}\label{subsec:su(3)}

The inclusion of baryon parity doublets can be done, as in Ref. \cite{PhysRevC.84.045208}, by considering the doubling of each baryon field in Eq. (\ref{baryonoctet}) of Appendix A, such that the mass term in a mean-field approximation reads
\begin{align}\label{intsym}
    \mathcal{L}_{mass}^{inv} = & -i\text{ m}_{0} \Tr \left(\bar{\Psi} \gamma_{5}\rho_{2} \Psi\right) + D_{s}^{(1)} \Tr \left(\bar{\Psi} \{\Sigma,\Psi \}  \right) + F_{s}^{(1)} \Tr \left(\bar{\Psi} \left[\Sigma,\Psi \right]  \right) + S_{s}^{(1)} \Tr \left( \Sigma \right) \Tr \left(\bar{\Psi} \Psi \right)\\ \nonumber
    &+ D_{s}^{(2)} \Tr \left(\bar{\Psi} \rho_{3} \{\Sigma,\Psi \}  \right) + F_{s}^{(2)} \Tr \left(\bar{\Psi} \rho_{3} \left[\Sigma,\Psi \right]  \right) + S_{s}^{(2)} \Tr \left( \Sigma \right) \Tr \left(\bar{\Psi} \rho_{3} \Psi \right) \, .
\end{align}
The mass matrix for each $\Psi_{ij}$ has nondiagonal terms that couple opposite parity states. These matrices are diagonalized by rotations such that
\begin{align}\label{transformedfield}\nonumber
\Psi^{'}_{ij} &= \frac{1}{\sqrt{2 \cosh{\delta_{ij}}}} \left(\begin{array}{cc} e^{\delta_{ij}/2} & e^{-\delta_{ij}/2} \gamma_{5} \\
e^{-\delta_{ij}/2} \gamma_{5} & e^{\delta_{ij}/2}
\end{array}\right) \Psi_{ij} \, , \\ 
\bar{\Psi}^{'}_{ij} &= \bar{\Psi}_{ij} \frac{1}{\sqrt{2 \cosh{\delta_{ij}}}} \left(\begin{array}{cc} e^{\delta_{ij}/2} & -e^{-\delta_{ij}/2} \gamma_{5} \\
-e^{-\delta_{ij}/2} \gamma_{5} & e^{\delta_{ij}/2}
\end{array}\right) \, ,
\end{align}
and each $\delta_{ij}$ is determined during the diagonalization procedure analogously to what was done for the SU(2) case (see Appendix A for details).

The masses $m_{i,p}^{inv}$ are given as functions of the nonstrange scalar condensate $\sigma$ and  the strange scalar condensate $\zeta \equiv \langle \zeta \rangle$ field as
\begin{subequations}
\begin{align}\label{massnucleon}
    m_{N, \pm}^{inv}=& \sqrt{(g^{(1)}_{N \sigma}\sigma + g^{(1)}_{N \zeta}\zeta)^2+ m_{0}^2} \pm g^{(2)}_{N \sigma} \sigma \pm g^{(2)}_{N \zeta} \zeta \, , \\  \label{masslambda}
    m_{\Lambda, \pm}^{inv}=& \sqrt{(g^{(1)}_{\Lambda \sigma}\sigma+g^{(1)}_{\Lambda \zeta}\zeta)^2+ m_{0}^2} \pm g^{(2)}_{\Lambda \sigma} \sigma \pm g^{(2)}_{\Lambda \zeta} \zeta \, , \\  \label{masssigma}
    m_{\Sigma, \pm}^{inv}=& \sqrt{(g^{(1)}_{\Sigma \sigma}\sigma+g^{(1)}_{\Sigma \zeta}\zeta)^2+ m_{0}^2} \pm g^{(2)}_{\Sigma \sigma} \sigma \pm g^{(2)}_{\Sigma \zeta} \zeta \, , \\ 
    m_{\Xi, \pm}^{inv}=& \sqrt{(g^{(1)}_{\Xi \sigma}\sigma + g^{(1)}_{\Xi \zeta}\zeta)^2+ m_{0}^2} \pm g^{(2)}_{\Xi \sigma} \sigma \pm g^{(2)}_{\Xi \zeta} \zeta \, .
    \label{masscascade}
\end{align}
\end{subequations}
The $16$ coupling constants thus introduced, $\{g^{(i)}_{N j}$, $g^{(i)}_{\Sigma j}$, $g^{(i)}_{\Lambda j}$, $g^{(i)}_{\Xi j}\}$ ($i=1,2$; $j=\sigma$,$\zeta$), are given as linear combinations the $6$ independent parameters $D_{s}^{(1)}$, $F_{s}^{(1)}$, $S_{s}^{(1)}$,$D_{s}^{(2)}$, $F_{s}^{(2)}$, $S_{s}^{(2)}$ present in Eq. (\ref{intsym}). If one wants to respect the structure imposed by chiral symmetry, one can choose $6$ of them freely. Choosing, as free parameters, $g^{(i)}_{N \sigma}$, $g^{(i)}_{\Sigma \sigma}$, $g^{(i)}_{\Lambda \sigma}$, the remaining coupling constants are given by

\begin{subequations}
\begin{align}
\label{explicitterms}
    g^{(i)}_{\Xi \sigma}=&\frac{1}{2}\left(-2 g^{(i)}_{N \sigma}+3 g^{(i)}_{\Lambda \sigma}+ g^{(i)}_{\Sigma \sigma} \right)\, , \\
    g^{(i)}_{N \zeta}=& \frac{\sqrt{2}}{4}\left(-4 g^{(i)}_{N \sigma}+3 g^{(i)}_{\Lambda \sigma}+ 2g^{(i)}_{\Sigma \sigma} \right)\, , \\    
g^{(i)}_{\Lambda \zeta}=& \frac{\sqrt{2}}{6}\left(-3 g^{(i)}_{\Lambda \sigma}+ 2g^{(i)}_{\Sigma \sigma} \right)\, ,\\
    g^{(i)}_{\Sigma \zeta}=& \frac{3\sqrt{2}}{2}g^{(i)}_{\Lambda \sigma}\, , \\
    g^{(i)}_{\Xi \zeta}=& \frac{\sqrt{2}}{4}\left(4 g^{(i)}_{N \sigma}-3 g^{(i)}_{\Lambda \sigma}\right)\, .
    \end{align}
\end{subequations}
%

\subsection{Mean-field Lagrangian density}\label{completeL}

In terms of mass and parity eingenstates, the full baryonic Lagrangian assumes the form
\begin{equation}
{\cal L}_B = \sum_{i, p=\pm}\bar{\psi}_{i,p}(i\gamma^\mu\partial_\mu- m_{i,p} +
\gamma^0\mu_{i}^{*} )\psi_{i,p}	\, ,
\end{equation}
where the index $i = N, \Lambda, \Sigma, \Xi $, runs through the baryon octet and the index $p$ sums over parity eigenstates. The baryon masses are denoted by $m_{i,p}$ and read
\begin{subequations}
\begin{align}
    m_{N, \pm}=& m_{N, \pm}^{inv} \, , \\ 
    m_{\Lambda, \pm}=& m_{\Lambda, \pm}^{inv} + (m_{1}+2m_{2})/3 \, ,\\ 
    m_{\Sigma, \pm}=& m_{\Sigma, \pm}^{inv} + m_{1} \, ,\\ 
    m_{\Xi, \pm}=& m_{\Xi, \pm}^{inv} + m_{1}+ m_{2} \, ,
\end{align}
\end{subequations}
where we also include explicit symmetry breaking terms in the hypercharge direction as in \cite{PhysRevC.57.2576} to improve on the vacuum value of baryon masses and hyperon potential depths:
\begin{equation}\label{intbre}
        \mathcal{L}_{\Delta m} \equiv -m_{1} \Tr \left(\Bar{\Psi}^{'} \Psi^{'} -\Bar{\Psi}^{'} \Psi^{'} S\right)-m_{2} \Tr \left( \Bar{\Psi}^{'} S \Psi^{'} \right)\, ,
\end{equation}
where $S = \text{diag} (0,0,1)$.

The effective chemical potentials are affected by the vector meson condensates. Assuming only the condensation of the fields $\langle \rho_{0}^{0} \rangle \equiv \rho$, $\langle \omega_{0} \rangle \equiv \omega$ and $\langle \phi_{0} \rangle \equiv \phi$, they read
\begin{subequations}
\begin{align}
\mu_{n/p}^*&=\mu_{n/p}-g_{N\omega}\omega-g_{N\phi}\phi\mp g_{N\rho}\rho \, , \label{muNs}\\[2ex]
\mu_{\Sigma^0}^*&=\mu_{\Sigma^0}-g_{\Sigma\omega}\omega-g_{\Sigma\phi}\phi \, , \\[2ex]
\mu_{\Sigma^\pm}^*&=\mu_{\Sigma^\pm}-g_{\Sigma\omega}\omega-g_{\Sigma\phi}\phi\pm g_{\Sigma\rho}\rho \, , \\[2ex]
\mu_{\Lambda}^*&=\mu_{\Lambda}-g_{\Lambda\omega}\omega-g_{\Lambda\phi}\phi \, , \\[2ex]
\mu_{\Xi^0/\Xi^-}^*&=\mu_{\Xi^0/\Xi^-}-g_{\Xi\omega}\omega-g_{\Xi\phi}\phi\pm g_{\Xi\rho}\rho \, . 
\end{align}
\end{subequations}
where only three of these couplings are free parameters while the rest are given by symmetry relations (see Appendix A). The mesonic part of the Lagrangian reads
\begin{equation}\label{mesonL} \allowdisplaybreaks
{\cal L}_M = \,\frac{1}{2}\partial_\mu\sigma\partial^\mu\sigma +\frac{1}{2}\partial_\mu\zeta\partial^\mu\zeta -\frac{1}{4}\omega_{\mu\nu}\omega^{\mu\nu}-\frac{1}{4}
 \phi_{\mu\nu}\phi^{\mu\nu}-\frac{1}{4}\rho_{\mu\nu}^0\rho^{\mu\nu}_0 - U(\sigma,\zeta)-V(\omega,\rho,\phi)	\, .	 			 
\end{equation}

In our approach, higher than the usual fourth-order linear sigma potential contributions are necessary to describe the vacuum and nuclear matter properties (see Sec. \ref{subsec:parameterfixing}), and to assure stability. Different relativistic mean field approaches \cite{Motohiro:2015taa,Mukherjee:2016nhb} also find that  higher-order contributions are necessary to  describe nuclear matter properties correctly. We make the assumption that $\zeta \equiv \sigma$, which is reasonable if we take into account that via PCAC we can relate the values of the scalar condensates to the decay constants of the pion and the kaon via:
\begin{equation}
    \sigma_{vac}=f_{\pi} \, , \qquad \zeta_{vac}=\frac{1}{\sqrt{2}}(2f_{K}-f_{\pi}).
\end{equation}
This can fulfill the condition $\zeta \equiv \sigma$ when considering $f_{K}$ around 112 MeV for pion decay constant of $f_{\pi}=93$ MeV, which aligns well with the experimental value of the kaon decay constant $f_{K} \simeq 110$ MeV \cite{Workman:2022ynf}. There are two practical reasons for using this approximation: first, for all the tested generalized potentials only an abnormal solution with $\sigma \simeq 0$ and $\zeta > \zeta_{VEV}$ was found as the global minima solution for all densities;  second, for performing most of the calculations we are concerned about, having only one order parameter is considerably simpler. With this assumption the chirally invariant piece of the vacuum potential, $U_{0}(\sigma, \zeta=\sigma) = U_{0}(\Phi)$, must be given as a function of the invariant scalar $\Phi = \frac{1}{2} \sigma^2$. We take the Taylor expansion around the vacuum value $\Phi$, which is done by a simple constant shift in the potential
\begin{equation}
U_{0}(\Phi) = \sum_{n=1}^4 \frac{a_n}{n!} 
(\Phi^2-\Phi_{0}^2)^n \, .
\end{equation}
Taking $\sigma_{vac} =f_{\pi}$ leads to
\begin{equation}\label{vacpotential}
U(\sigma) = \sum_{n=1}^4 \frac{a_n}{n!} 
\frac{(\sigma^2-f_\pi^2)^n}{2^n}-m_{\pi}^{2}f_{\pi}(\sigma-f_\pi) \, ,
\end{equation}
where the parameter $a_1=m_{\pi}^{2}$  guarantees the correct pion vacuum mass, $m_{\pi}= 139$ MeV, and an explicit symmetry breaking term was added to ensure that $U(\sigma)$ has a minimum at $\sigma = f_{\pi}$, with the pion decay constant set to $f_{\pi} = 93 $ MeV. In this approach, the mass of the chiral condensate is given by $m_{\sigma}^2=m_{\pi}^2 +f_{\pi}^2 a_{2}$ which for the range of values that the parameter $a_{2}$ assumes in our results yields $m_{\sigma} \sim (500-700)$ MeV. The vacuum potential in (\ref{vacpotential}) is the same adopted previously for two-flavors \cite{Fraga:2018cvr,Schmitt:2020tac,Drews:2013hha,Drews:2014spa} and also in Ref. \cite{Fraga:2022yls}, where hyperonization was considered.

The vector meson potential in (\ref{mesonL}) reads
\begin{equation} \allowdisplaybreaks
V(\omega,\rho,\phi) = \,\frac{m_\omega^2}{2}\omega^2 + \frac{m_\rho^2}{2}\rho^2+ \frac{m_\phi^2}{2}\phi^2 + V_{quartic}(\omega,\rho,\phi),	 
 \label{eq:Lmeson}			 
\end{equation}
where for the vector meson masses we take $m_{\omega} = 782$ MeV, $m_{\rho} = 775$ MeV and $m_{\phi} = 1020$ MeV. The last term  is the self interaction that introduces quartic contributions, $\sim\omega^4$, to the vector interactions which are responsible for the high density behavior of $c^2_{s}=1/3$, as was explicitly demonstrated for isospin symmetric matter \cite{Fraga:2022yls}. Moreover, the coupling  $\sim \omega^2\rho^2$ yields a better agreement with the experimental data on the value of the slope parameter $L$ (for a recent overview of the various estimates for $L$, see Ref. \cite{Sotani:2022hhq}). The usual parametrization of the matrix containing the vector meson fields is
\begin{eqnarray}
V_\mu & \equiv & \frac{1}{\sqrt{2}}\left(\begin{array}{ccc} \displaystyle{\frac{\rho^0_\mu}{\sqrt{2}} + \frac{\omega_\mu}{\sqrt{2}}} & \rho^+_\mu & K^{*+}_\mu \\ \rho^-_\mu &
\displaystyle{-\frac{\rho^0_\mu}{\sqrt{2}} + \frac{\omega_\mu}{\sqrt{2}}}  & K^{*0}_\mu  \\[3ex] K^{*-}_\mu & \bar{K}^{*0}_\mu & \phi_\mu 
\end{array}\right) \, . 
\end{eqnarray}

There are two 
SU(3) symmetric structures for the vector self-interactions at fourth order in the vector fields:
\begin{align}
V_{quartic}(\omega,\rho,\phi) = & d_1(\Tr[V_{\mu}V^{\mu}])^2 + d_2\Tr[(V_{\mu}V^{\mu})^2]\,  \\ \nonumber
=& \frac{d_1}{4}(\omega^{2}+\rho^{2}+\phi^{2})^2 + \frac{d_2}{8}\left(\omega^4+\rho^4+6 \omega^{2}\rho^{2}\right) \, ,
\end{align}
where in the last line we kept only the fields assumed to condense. In the following we set $d_1=0$ and denote $d=d_2$ in line of the discussion above.

Finally, noninteracting electrons and muons are added to the system via
\begin{equation}
 \mathcal{L}_\text{leptons}=\sum_{i=e,\mu}  \bar{\psi}_{i}(i\gamma^{\mu}\partial_{\mu}-m_{i}+
\gamma^{0}\mu_{i})\psi_{i} \, ,
\end{equation}
where lepton masses are taken to be $m_{e}=0.5$ MeV and $m_{\mu}=106$ MeV. In this work we assume that neutrinos have mean free paths larger than the size of the system, which in our case is the typical size of a cold neutron star ($\sim 10$ km). For this reason, they will be omitted throughout the rest of the text. 

\subsection{Free energy and stationary equations}\label{subsec:freeenergy}

In the no-sea approximation (fermionic vacuum contributions ignored), the standard field theory computation of the zero-temperature free energy density leads to
\begin{equation}\label{energyden}
\Omega = -\left(\nabla \omega\right)^2-\left(\nabla \rho \right)^2-\left(\nabla \phi \right)^2+\left(\nabla \sigma\right)^2+ \frac{\left(\nabla \sigma\right)^2}{2 e^2} + \Omega_{hom} \, ,
\end{equation}
with $\Omega_{hom}$ defined as
\begin{equation}\label{hompot}
\Omega_{hom} = V(\omega,\rho,\phi) + U(\sigma) -\sum_{i,p=\pm} p(\mu_{i}^{*},m_{i,p}^{*})-p(\mu_{e},m_{e})-p(\mu_{\mu},m_{\mu}) \, ,
\end{equation}
which depends only on the condensates themselves, not their derivatives, being the part of the potential that determines the homogeneous solution. The electric charge is denoted $e$ and, in Heaviside-Lorentz units, has a value of $e\simeq 0.3$. In Eq. (\ref{hompot}) the partial pressures have the form
\begin{equation}
    p(\mu,m)= \frac{\Theta(\mu,m)}{8 \pi ^2}\left[\left( \frac{2}{3}k_{F}^3-m^{2}k_{F}\right)\mu +m^{4} \ln\left(\frac{k_{F}+\mu}{m} \right)\right]\, ,
\end{equation}
with the Fermi momentum
$k_{F}=\sqrt{\mu^{2}-m^2}$.

The chemical potentials for each baryon are not independent, but in three-color QCD they are fundamentally related to the three chemical potentials of $u$, $d$ and $s$ quarks. The condition of electroweak equilibrium makes it possible to narrow this freedom down to two chemical potentials: $\mu_{B}$, related to the baryon number density, and $\mu_{e}$ ($\mu_{\mu}=\mu_{e}$, in beta equilibrium), related to the lepton number density that will affect only the charged baryons, such that:
\begin{subequations}
\begin{align}
\mu_{n}&=\mu_{\Sigma^0}=\mu_{\Lambda}=\mu_{\Xi^0}=\mu_{B} \, , \label{muNtrue}\\[2ex]
\mu_{p}&=\mu_{\Sigma^+}=\mu_{B}-\mu_{e} \, , \\[2ex]
\mu_{\Sigma^-}&=\mu_{\Xi^-}=\mu_{B}+\mu_{e} \, .
\end{align}
\end{subequations}
independent of the parity of the baryonic state. From the free energy density of Eq. (\ref{energyden}), we derive the following Euler-Lagrange equations
\begin{subequations} \label{stats}
\begin{eqnarray}
\nabla^2 \sigma &=&\frac{\partial \Omega}{\partial\sigma} = \frac{\partial U}{\partial \sigma} +\sum_{i, p=\pm} g_{i\sigma} n_{si,p} \, , \label{Eqsig}\\[2ex]
-\nabla^2 \omega &=&\frac{\partial \Omega}{\partial\omega} = \frac{\partial V}{\partial \omega} +\sum_{i, p=\pm} g_{i\omega} n_{i,p} \, , \label{Eqw}\\[2ex]
-\nabla^2 \phi &=&\frac{\partial \Omega}{\partial\phi} = \frac{\partial V}{\partial \phi} +\sum_{i, p=\pm} g_{i\phi} n_{i,p} \, , \label{Eqphi}\\[2ex]
-\nabla^2 \rho &=&\frac{\partial \Omega}{\partial\rho} = \frac{\partial V}{\partial \rho} +\sum_{i, p=\pm} g_{i\rho} n_{i,p} \, , \label{Eqrho}
\end{eqnarray}
\end{subequations}
where
\begin{equation}
n_{i,p}\equiv n(\mu_i^*,m_{i,p}^{*}) \, , \qquad n_{si,p}\equiv n_s(\mu_i^*,m_{i,p}^{*}) \, ,
\end{equation}
with
\begin{subequations}
\begin{eqnarray}
n_s(\mu,m) &=& -\frac{\partial p}{\partial m} = \Theta(\mu-m)\frac{m}{\pi^2}
\int_0^{k_F}\frac{k^2 dk} {\sqrt{k^2+m^2}} = \Theta(\mu-m)\frac{m}{2\pi^2}\left(k_F\mu-m^2\ln\frac{k_F+\mu}{m}\right) \, , \\[2ex]
n(\mu,m) &=& \frac{\partial p}{\partial \mu} = \Theta(\mu-m)\frac{k_F^3}{3\pi^2} \, .
\end{eqnarray}
\end{subequations}

We can also derive a Euler-Lagrange equation for $\mu_{e}$
\begin{equation}
\frac{\nabla^2 \mu_{e}}{e^2} = \frac{\partial\Omega}{\partial\mu_e} = -n_{p_{+}}-n_{p_{-}}-n_{\Sigma^{+}_{+}}-n_{\Sigma^{+}_{-}}+n_{\Sigma^{-}_{+}}+n_{\Sigma^{-}_{-}}+n_{\Xi^{-}_{+}}+n_{\Xi^{-}_{-}}+n_e+n_\mu \, ,
\end{equation}
such that local electric charge neutrality corresponds to $\nabla^2 \mu_{e} = 0$.

\subsection{Parameter fixing}\label{subsec:parameterfixing}

In this section we explain how the free parameters of the model are fixed to reproduce properties of the vacuum and of isospin symmetric nuclear matter at saturation (see Appendix B for more details). The parameters $g_{N \sigma}^{(2)}$, $g_{\Sigma\sigma}^{(2)}$ and
$g_{\Lambda\sigma}^{(2)}$ are fixed by the vacuum mass splitting between parity partners for $N,\Lambda$ and $\Sigma$ baryons (Eqs. (\ref{massnucleon}), (\ref{masslambda}) and (\ref{masssigma}), for $\zeta=\sigma=f_\pi$). The mass of the negative parity partners are usually assumed to be $m_{N,-}^{\rm vac} = 1535\, {\rm MeV}$, $m_{\Sigma,-}^{\rm vac} = 1750\, {\rm MeV}$ and $m_{\Lambda,-}^{\rm vac} = 1670\, {\rm MeV}$. Other assignments are also possible since the experimental data shows multiple resonances that could potentially be identified to the chiral partners of the baryon in the baryon octet \cite{Workman:2022ynf,proceedings2019013007}.

One still has to fix the values of $12$ other parameters: $a_2$, $a_3$, $a_4$, $d$, $g_{N\omega}$, $g_{N\rho}$, $g_{N \phi}$, $g_{N \sigma}^{(1)}$, $g_{\Sigma \sigma}^{(1)}$, $g_{\Lambda \sigma}^{(1)}$, $m_{1}$, $m_{2}$. This set of parameters will be fixed with the help of $9$ properties of isospin-symmetric nuclear matter at saturation and the $3$ vacuum masses $m_{N,+}^{\rm vac} = 939\, {\rm MeV}$, $m_{\Sigma,+}^{\rm vac} = 1190\, {\rm MeV}$ and $m_{\Lambda,+}^{\rm vac} = 1116\, {\rm MeV}$. The nuclear matter properties to be considered are: the binding energy $E_B=16\, {\rm MeV}$, saturation density $n_0=0.15\, {\rm fm}^{-3}$, symmetry energy $S=30\, {\rm MeV}$, slope of the symmetry energy $L\simeq (40-70)\, {\rm MeV}$, effective nucleon mass at saturation $M_0\simeq (0.55-0.75) \, m_{N,+}^{\rm vac}$, incompressibility at saturation $K = 260\, {\rm MeV}$, and the hyperon potential depth $U_{\Lambda} = -30\, {\rm MeV}$. We also fix $U_{\Sigma} = 20$ MeV and $U_{\Xi} = -20$ MeV. We checked that mild modifications are found if we instead consider other values for hyperon potentials within $U_{\Sigma} \in \left[0, 30\right]$ MeV and $U_{\Xi}  \in \left[-30,0\right]$ MeV. These values have been selected by considering experimental evidence that makes reasonable placing the $U_{\Lambda}$ value near -28 MeV and the $U_{\Xi}$ values near -20 MeV (albeit with less certainty). Although the $U_{\Sigma}$ values remain uncertain, it is noteworthy that $\Sigma$ baryons are absent in hypernuclei bound states. This observation provides support to the notion of a repulsive, positively valued optical potential. For recent research and comprehensive discussions regarding hypernuclei physics and its relevance to neutron star physics, see Refs. \cite{Gal:2016boi,Haidenbauer:2019boi,Tolos:2020aln,Haidenbauer:2018gvg,Gal:2019llv,Friedman:2022bju,Friedman:2023ucs,Friedman:2022bor}. The rest of the couplings are all determined by the SU(3) relations (\ref{explicitterms}) and (\ref{vectorcoupling}). In our approach the masses of the cascade baryon $m_{\Xi,+}^{\rm vac}$ and its parity counterpart $m_{\Xi,-}^{\rm vac}$ are both determined after the parameter fixing. For the parameters adopted here,  $m_{\Xi,+}^{\rm vac} \sim 1330$ MeV (see Table \ref{table:para}) which is in good agreement with usually adopted mean masses of $1318$ MeV. The value of $m_{\Xi,-}^{\rm vac}$ in our approach is $m_{\Xi,-}^{\rm vac} \sim 1845$ MeV. 

At high densities, the low-energy properties of matter provide limited information, and it is unreasonable to expect that the relevant degrees of freedom are the same as those in confined matter. At some point, partonic degrees of freedom, as predicted by perturbative QCD, become relevant. To overcome the lack of experimental microphysics input at high densities, we incorporate constraints from astronomical observations, which significantly restrict the high-energy portion of the equation of state derived from our model (more on this matter in the next section). Furthermore, we enforce the condition that, at asymptotically high densities, the speed of sound approaches the conformal limit of $c_{s}^2 \rightarrow 1/3$. Within our model, this is achieved through the quartic self-coupling of the vector meson.

\section{Results}\label{sec:results}

\subsection{Parameter space}

We narrow down our parameter space by selecting values that allow for the existence of stable star configurations with masses of at least 2.0 solar masses. We solve the Tolman-Oppenheimer-Volkoff equations using as input the equation of state (EoS) obtained via the usual thermodynamic relation $\epsilon = -P +\mu_{B} n_{B}$ for densities higher than 1.1 $n_{0}$. For lower densities, we adopt the EoS proposed by Hebeler et al. \cite{Hebeler:2013nza}. Although the use of this low-density chiral effective theory EoS has a negligible impact on the maximum mass of neutron stars, it provides significant corrections to the radii, particularly for the less massive stars \cite{Kalaitzis:2019dqc,Perot:2019gwl,Perot:2020gux,Margaritis:2021gdr}. 
Therefore, it is reasonable to impose additional constraints on the parameter space by considering recent constraints on the radii of neutron stars. We specifically select parameters that lead to neutron star families falling within the one-sigma error bars of the measured radii of 1.4 \(\textup{M}_\odot\) and 2.0 \(\textup{M}_\odot\) stars consistent with the analyses conducted in Refs. \cite{Miller:2019cac, Riley:2019yda, Miller:2021qha, Riley:2021pdl} using data from NICER on the pulsars PSR J0030+0451 and PSR J0740+6620. This approach allows us to identify the different qualitative characteristics that emerge from the possible parameter choices within reasonable experimental and observational constraints. In our analysis, we primarily focus on varying key properties, namely the effective mass at saturation $M_{0}$, the invariant mass $m_{0}$ and the slope parameter $L$. These parameters have a significant impact on our results and are responsible for the most pronounced changes. For simplicity, we only vary these parameters as minor modifications within their experimental uncertainty range. The remaining nuclear properties yield negligible changes in our final conclusions.\par

\begin{figure}
\begin{center}
\includegraphics[width=0.32\textwidth]{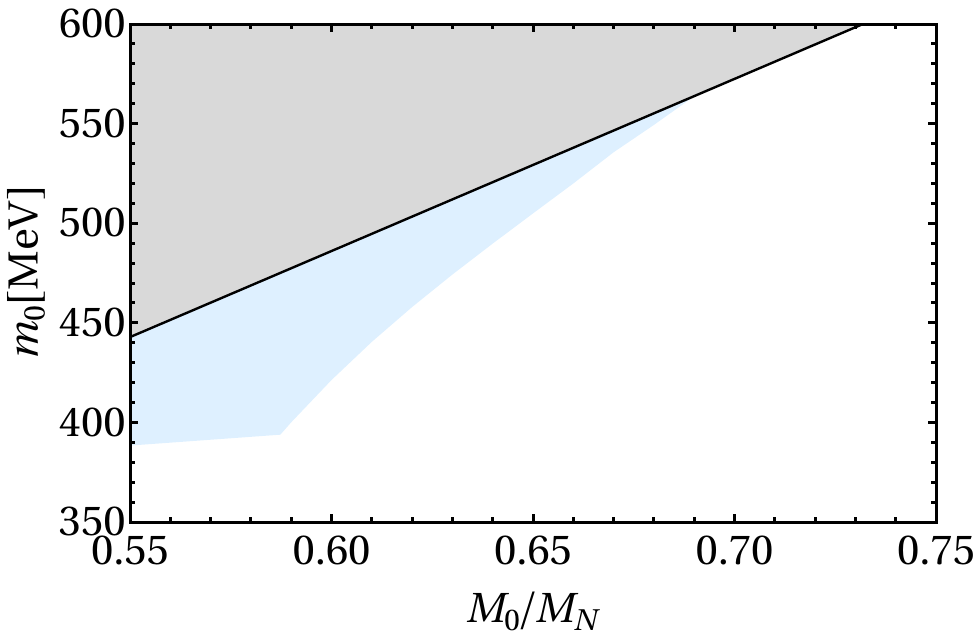}
\includegraphics[width=0.32\textwidth]{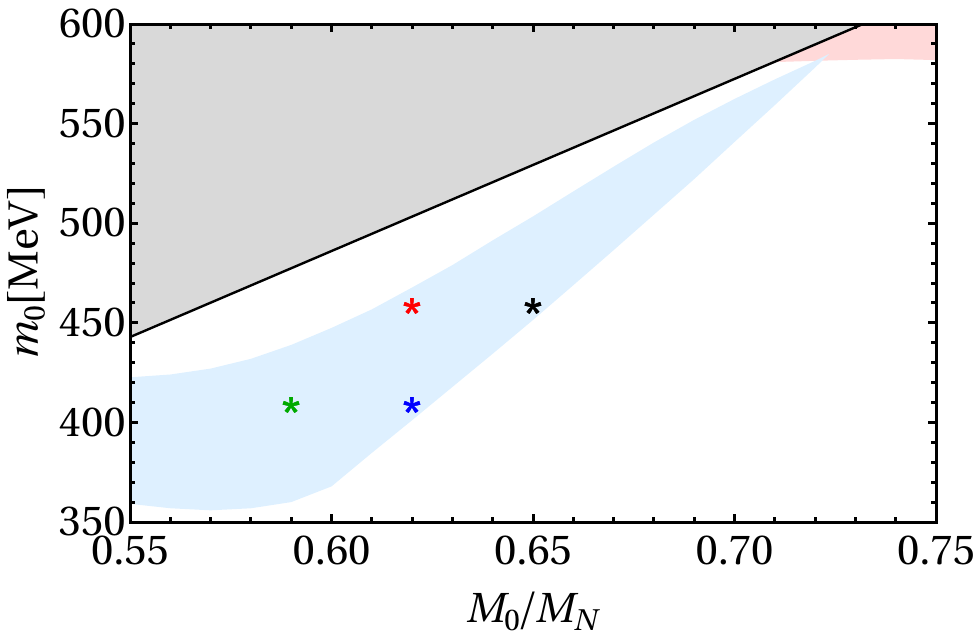}
\includegraphics[width=0.32\textwidth]{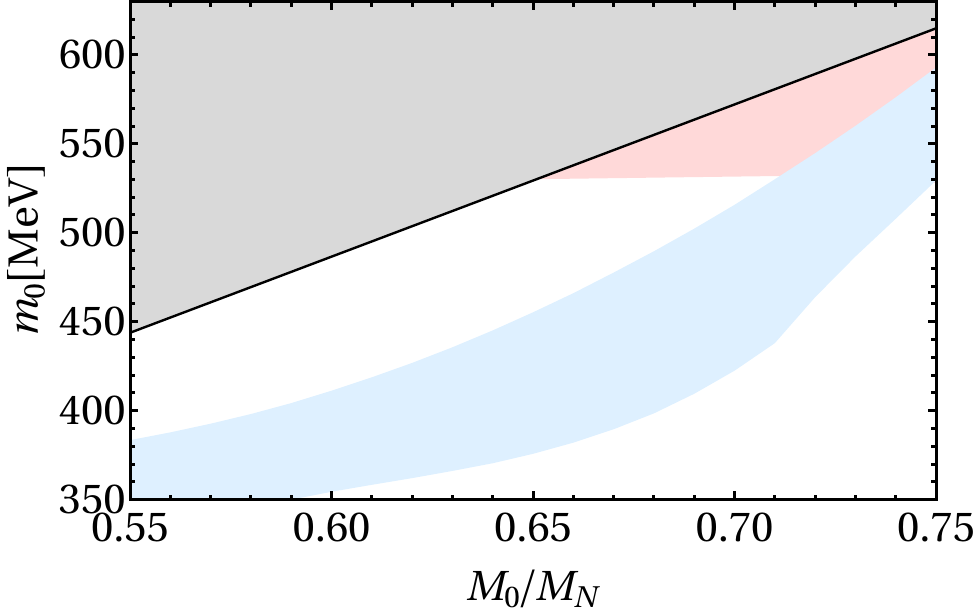}
\caption{From left to right the parameter space for $L=65, 70$ and $80$ MeV. In all panels the blue region corresponds to parameters that satisfy the astrophysical constraints, the red region corresponds to parameters that lead to crossovers. The gray region leads to a unstable vacuum potential with $a_{4}<0$. The asterisks mark the sample sets of parameters of Table \ref{table:para}.}
\label{fig1}
\end{center}
\end{figure}

For a given baryon chemical potential, we solve the stationary equations [ignoring the Laplacian contribution in Eqs. (\ref{stats})] for homogeneous meson condensates, incorporating the equation of local charge neutrality and enforcing beta equilibrium.  Figure \ref{fig1} displays the $M_{0} \times m_{0}$ parameter space for three values of $L=65,70$ and $80$ MeV, respectively. The parameter space is limited from above by the condition that the potential 
$U(\sigma)$ remains bounded, which translates to the requirement that parameter $a_4$ in Eq. (\ref{vacpotential}) is greater than zero. The excluded region are shaded in gray. The blue region corresponds to parameters that lead to star masses and radii within the astrophysical constraints. The red region represents parameter values for which the chiral restoration occurs via a crossover.

A crossover can be attained for the highest values of effective mass $M_{0}$ and invariant mass parameter $m_{0}$, being restricted to the upper right part of the parameter space in all cases. Its intersection with the astrophysical region becomes smaller for lower values of the slope parameter $L$. For $L< 70$ MeV, and therefore toward more realistic values of the slope parameter, the astrophysical constraints become incompatible with a crossover. We also observe that lower values of $L$ tend to drag the blue region toward lower values of $M_{0}$ and higher values of $m_0$ for each given $M_{0}$.

To exemplify the different ways that chiral restoration can take place in the model, we take four points in the allowed parameter space, corresponding to the asterisks in the middle panel of Fig. \ref{fig1}, with $L= 70$ MeV, and with values of $M_{0}$ and $m_{0}$ in Table \ref{table:para}. We also show the critical chemical potentials for the first-order phase transitions and the potential for the onset of strangeness, $\mu_{c}$ and $\mu_{S}$, respectively. In all cases, the negative parity baryons just onset at the (approximate) chiral restored phase being present already just after the transition, which implies that the density of negative parity partners serves as a order parameter for the transition, being zero at the chirally broken phase and nonzero at the chirally restored phase. 
We also provide the corresponding $m_1$ and $m_2$ parameter values related to explicit symmetry breaking in the hypercharge direction introduced in Eq. (\ref{intbre}). The significant $m_1$ values indicate a considerable strange quark bare mass. As our focus is on understanding the broad phenomenological effects of the interplay between parity doubling and hyperonization, we anticipate our outcome to be only qualitatively similar to a more realistic model with a realistic strange quark mass ($m_{s} =95$ MeV). In this context, we prioritize the accurate representation of nuclear properties, such as baryon masses and hyperon potential depths, rather than emphasizing the reproduction of bare quark masses, which are not even treated as degrees of freedom in our approach.

\begin{table}
\begin{tabular}{c |c | c || c | c | c | c | c | c | c | c | c | c | c | c } 
 \hline
    Sets  & $\;$$M_0/M_N$$\;$ & $m_0 [{\rm MeV}]$ & $g_{N \sigma}$ & $g_{\Sigma \sigma}$ & $g_{\Lambda \sigma}$ & $g_{N \omega}$ & $g_{N \phi}$ & $g_{N \rho}$ & $\mu_{c} [{\rm MeV}]$ & $\mu_{S} [{\rm MeV}]$ & $m_{\Xi,+} [{\rm MeV}]$  & $m_{\Xi,-} [{\rm MeV}]$ & $m_{1} [{\rm MeV}]$  & $m_{2} [{\rm MeV}]$ \\ [0.5ex]
 \hline\hline
  \textcolor{darkgreen}{Green} & 0.59 & 410  & 12.5 & 9.4 & 11.9 & 13.3 & 5.9 & 4.5 & 1050.6 & 1402.3 & 1330.2 & 1845.2 & 506.7 & 62.7\\ 
 \hline
  \textcolor{blue}{Blue} & 0.62 & 410  & 12.5 & 9.2 & 11.9 & 12.5 & 5.3 & 4.4 & 1047.4 & 1421.2 & 1330.3 & 1845.3 & 521.2 & 62.5 \\
 \hline
  \textcolor{red}{Red} & 0.62 & 460 & 12.3 & 9.4 & 11.8 & 11.5 & 5.8 & 3.6 & 1097.1 & 1402.9 & 1330.4 & 1845.4 & 496.4 & 68.9\\
 \hline
  Black & 0.65 & 460 & 12.3 & 9.0 & 11.6 & 11.8 & 4.8 & 4.4 & 1093.5 & 1419.5 & 1330.5 & 1845.5 & 512.6 & 67.0\\
 \hline
\end{tabular}
\caption{Values of couplings, critical potential of the phase transition $\mu_{c}$, strangeness onset potential $\mu_{S}$ and masses $m_{\Xi,+}$, $m_{\Xi,-}$, $m_{1}$, $m_{2}$ for four different choices of Dirac mass at saturation
$M_{0}$ and the invariant mass $m_{0}$. In all cases, the slope parameter L = 70 MeV, while K = 260 MeV and S = 30 MeV are fixed. The Hyperon potential depths are fixed to $U_{\Sigma}=20$ MeV and $U_{\Xi}=-20$ MeV.}
\label{table:para}
\end{table}

\begin{figure}
\begin{center}
\includegraphics[width=0.65\textwidth]{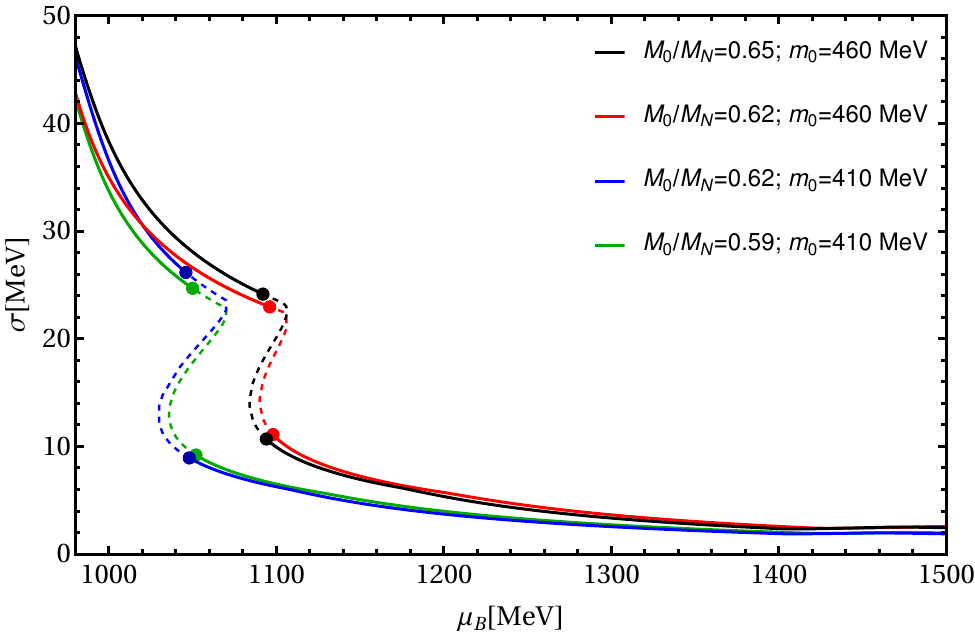}
\caption{Scalar condensate as a function of baryon chemical potential for the four parameter choices of Table \ref{table:para}.  Solid (Dashed) lines correspond to stable (metastable) solutions in each case.}
\label{fig:tunability}
\end{center}
\end{figure}

Figure \ref{fig:tunability} illustrates the behavior of the scalar condensate as a function of the chemical potential for $\mu_{B} \geq 980$ MeV, which corresponds to a baryonic density of 1.1 $n_{0}$ in all cases. The figure includes stable (solid lines) and spinodal curves (dashed lines). The turning points where the first-order transition occurs are marked with dots. In all cases the high density phase is just approximately chirally symmetric, approaching $\sigma =0$ only for very high chemical potentials, which is expected since there are explicit chiral symmetry breaking terms in the Lagrangian.

A remarkable aspect of the model presented in Ref. \cite{Fraga:2022yls}, which did not consider parity doublets, is its implication for the formation of massive stars. In that context, stars with 2.0 solar masses can only be formed when a robust and early first-order phase transition takes place, resulting in stars predominantly composed of chirally restored matter. However, with the inclusion of negative parity baryons, the chiral phase transition, while remaining first-order, gains the flexibility to manifest in different regions of the phase diagram. When examining curves that share the same invariant mass and yet exhibit different effective nucleon masses, it becomes evident that the main parameter influencing the critical potential for the phase transition is the magnitude of the chiral invariant mass $m_{0}$, assuming larger values for larger values of $m_{0}$. Furthermore, variations in the nucleon effective Dirac mass $M_{0}$ appear to directly impact the jump in the order parameter $\Delta \sigma$ during the phase transition, with $\Delta \sigma$ consistently increasing as $M_{0}$ increases. This observation suggests that adjusting the values of the invariant mass within the blue region of Fig. \ref{fig1} allows for precise tuning of both the position and the magnitude of the phase transition. This ``tuning" feature can be used to explore different phase transition scenarios within the same model, all consistent with the same experimental constraints.

The invariant mass, represented by $m_{0}$, plays a pivotal role in quantifying the contribution of chiral symmetry breaking to the generation of baryon mass. A higher invariant mass signifies that the chirally restored phase primarily consists of massive baryons. On the other hand, as $m_{0} \rightarrow 0$, the chirally restored phase becomes dominated by lighter, almost massless baryons. In a broader perspective, $m_{0}$ may arise through the condensation of scalar fields, such as the glueball condensate \cite{Heinz:2013hza,PhysRevD.82.014004}. If the theory exhibits dilatation invariance, we would anticipate $m_{0}$ to vary with density, decreasing as density increases. In our present context, our approach can be viewed as a ``frozen glueball" description, where we neglect any density dependence of the glueball condensate. The running of $m_{0}$ and its connection to the trace anomaly of QCD and deconfinement will be postponed to a later work. With this in mind, our focus in the subsequent section of this study will solely be on the value of $M_{0}/M_{N} = 0.65$, corresponding to the black set of parameters.

\subsection{Surface tension}\label{sec:surfacetension}
\subsubsection{Domain wall}
The surface tension is determined by the classical configuration that connects different homogeneous phases \cite{doi:10.1063/1.1730447,Langer:1969bc,LANGER197461,Coleman:1977py,Callan:1977pt,Linde:1980tt,Linde:1981zj,domb1983phase,Fraga:2018cvr,Schmitt:2020tac,Fraga:2022yls}, as specified by Eqs. (\ref{stats}). In the specific case of domain walls, the surface tension is well-defined as the energy difference per unit area between the configuration of the domain wall and the homogeneous configuration of each phase. Essentially, the surface tension represents the energy cost associated with transitioning from the first minimum to the second minimum by following a path in configuration space that satisfies the equations of motion. The domain wall solution can be seen as the critical bubble solution when its radius approaches infinity, particularly applicable in the immediate vicinity of the phase transition.

In the domain wall geometry, the Euler Lagrange equations become one-dimensional. Thus, one can make the simplification $\nabla^2 \rightarrow d^2/dx^2$ in Eqs. (\ref{stats}). To solve the system of differential equations at the phase transition, we employ the numerical method of over-relaxation with boundary conditions $\textbf{b}{\pm} (x=\pm \infty)$, where $\textbf{b}=(\sigma,\omega,\rho,\phi,\mu{e})$. At first glance, one might consider using the degenerate solutions at the phase transition as boundary values. However, a subtle issue arises due to the disparity in lepton chemical potential between the phases connected via a Maxwell construction, despite having identical pressure. Interpolating between these lepton chemical potential values would imply the presence of a nonhomogeneous lepton background across the interface region, which contradicts expectations during nucleation timescales that should be much smaller than the typical timescales associated with electromagnetic forces. To address this problem, we set $\mu_{e}$ to its value in the chiral broken phase $\mu_{e, \chi \text{Broken}}$. This decision is based on our primary focus being the nucleation process of the chirally restored phase within the predominantly chirally broken phase. Subsequently, we search for the chirally restored solution that possesses the same lepton chemical potential. Although this adjustment slightly modifies the critical value of the baryon chemical potential for the phase transition, it ensures that, when calculating the surface tension, we interpolate between condensate solutions embedded in a homogeneous lepton background.

Once the numerical solution is obtained, the surface tension is computed as
\begin{equation}
    \Sigma = \int_{-\infty}^{+\infty} dx \left[\frac{1}{2}( \sigma^{' 2}(x)-\omega^{' 2}(x)-\rho^{' 2}(x)-\phi^{' 2}(x)) + \Omega(x)-\Omega(\pm \infty)\right] \, .
\end{equation}


\begin{figure}
\begin{center}
\includegraphics[width=0.65\textwidth]{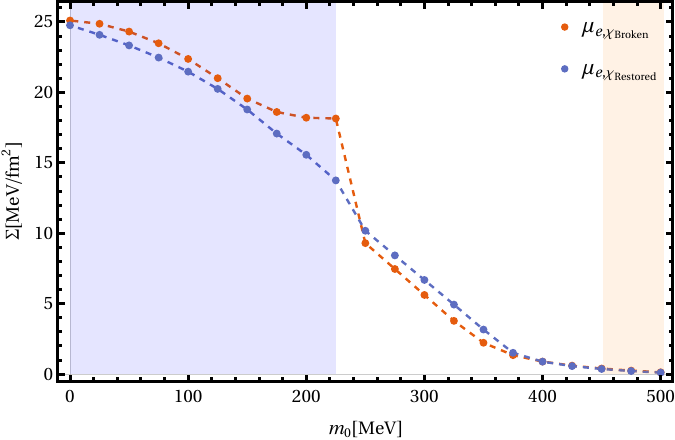}
\caption{The surface tension as a function of $m_{0}$ in the domain wall setup for both lepton chemical potentials at the phase transition. The region shaded in yellow corresponds to the values of the invariant mass that satisfy astrophysical constraints. On the other hand, the region shaded in blue indicates values of $m_{0}$ for which nuclear matter exists solely as a metastable solution and the chirally broken phase used in the surface tension calculation is the vacuum phase.}
\label{fig:surfacetension}
\end{center}
\end{figure}

To evaluate the error in the choice of lepton chemical potentials, we also conduct the same calculation using the lepton chemical potential of the chirally restored phase $\mu_{e, \chi \text{Restored}}$. Figure \ref{fig:surfacetension} exhibits the surface tension as a function of the invariant mass for both charge chemical potentials. The yellow-shaded region represents the invariant mass values that meet astrophysical constraints, while the blue-shaded region corresponds to values of $m_{0}$ that lead to nuclear matter existing solely as a metastable solution. For the case of metastable nuclear matter, the surface tension is computed by connecting the vacuum phase, where all baryons have their vacuum masses, directly to the chirally restored phase. Both calculated surface tensions exhibit similar values when $m_0$ assumes essentially the same magnitude for $m_0 > 400 \, \text{MeV}$, with the chirally broken chemical potential yielding slightly higher values for $\Sigma$. In the region of $225 \, \text{MeV} < m_0 < 400 \, \text{MeV}$, the $\mu_{e, \chi \text{Restored}}$ results in a higher surface tension value relative to the values associated with $\mu_{e, \chi \text{Broken}}$. In the bluish region, the behavior is reversed, with the chirally broken chemical potential leading to larger values of $\Sigma$ after a jump occurring at the point where nuclear matter ceases to be stable. This jump is associated with a corresponding leap in the values of the lepton chemical potential, shifting from a finite value for $m_0 > 225 \, \text{MeV}$ to exactly $\mu_e = 0$, which represents the vacuum value. Interestingly, achieving physical masses and radii for neutron stars requires very low values of surface tension. These values are approximately an order of magnitude smaller than those previously calculated in works such as \cite{Palhares:2010be,Pinto:2012aq,Mintz:2012mz,Lugones:2013ema,Fraga:2018cvr, Schmitt:2020tac}, where no parity doubling was considered. The reduced surface tension values directly stem from the fact that the energy density discontinuity connecting the phases is much less pronounced when the baryonic species remain massive in the chirally restored phase, which explains the monotonic decreasing behavior of its value with respect to the invariant mass.

\subsubsection{Bubbles}

To estimate the behavior of the surface tension in the spinodal regions, where the nucleating critical bubble assumes a finite size, we assume spherical symmetry \cite{doi:10.1063/1.1730447,Langer:1969bc,LANGER197461,Coleman:1977py,Callan:1977pt,Linde:1980tt,Linde:1981zj,domb1983phase}. The critical bubble profile is obtained by solving Eqs. (\ref{stats}) with $\nabla^2 \rightarrow d^2/dr^2 + \frac{2}{r} d/dr$ for the specific boundary conditions: $\textbf{b}(+\infty) = (\sigma_{\infty},\omega_{\infty},\rho_{\infty},\phi_{\infty})$, along with the condition $\frac{d\textbf{b}}{dr}|_{r=0}= \textbf{0}$. Here, $\textbf{b}(+\infty)$ represents the solution of the homogeneous equations corresponding to the metastable phase, which have a higher free energy. The values of the condensates at the center of the bubble are dynamically determined via a shooting method algorithm.

\begin{figure}
\begin{center}
\includegraphics[width=0.65\textwidth]{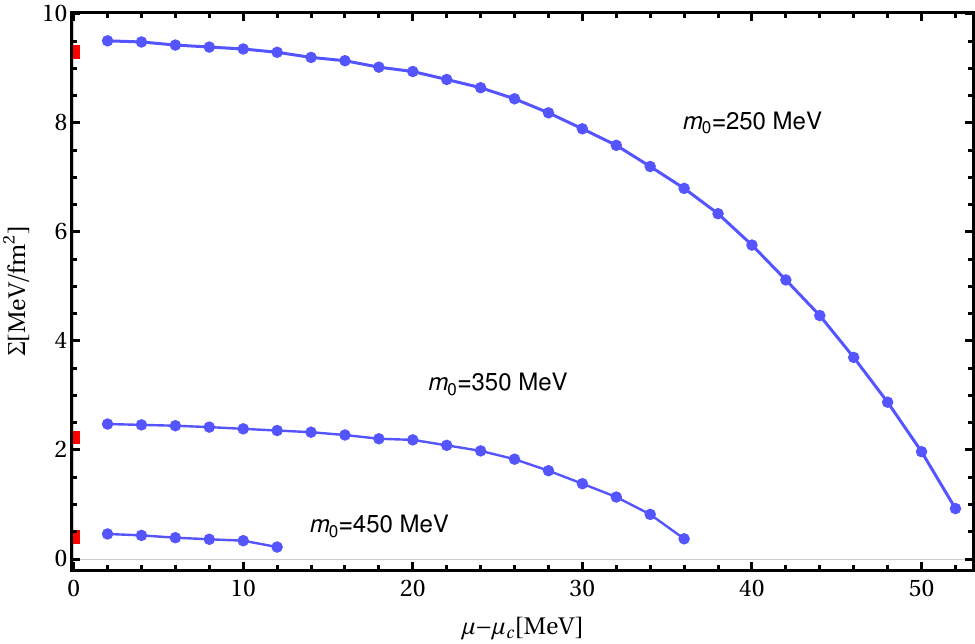}
\caption{Surface tension of bubbles in the spinodal region for $m_{0} = 250, 350, 450$ MeV. Red squares mark the corresponding domain wall solution at the phase transition.}
\label{fig:bubbles}
\end{center}
\end{figure}

To calculate the bubble profile for a single field, the standard semiclassical approach \cite{Coleman:1977py} is mathematically equivalent to the problem of a point particle moving along a potential valley under the influence of a time-dependent viscous force\footnote{Notice that one is lead to the euclidean space and the ``Newtonian dynamics'' happens in the inverted potential \cite{Coleman:1977py}.}. To leverage this approach, we utilize the ``one-condensate approximation" described in Ref. \cite{Fraga:2018cvr}. This approximation involves neglecting spatial derivatives in Eqs. (\ref{Eqw},\ref{Eqphi},\ref{Eqrho}), allowing us to solve these three equations for $(\omega(\sigma), \rho(\sigma), \phi(\sigma))$ for a given $\sigma$ value. Consequently, the Euler-Lagrange equation for the $\sigma$ condensate simplifies to a straightforward second-order differential equation:
\begin{equation}
    \frac{d^2 \sigma}{dr^2} + \frac{2}{r} \frac{d\sigma}{dr} =\frac{\partial \tilde{\Omega}}{\partial\sigma} \, ,
\end{equation}
where $\tilde{\Omega} \equiv \Omega (\sigma,\omega(\sigma),\rho(\sigma),\phi(\sigma),\mu_{e})$ represents a function that depends solely on the $\sigma$ condensate. We fix the lepton chemical potential at its values within the chirally broken, metastable phase in the same fashion performed previously, in the case of the domain wall.

In the one-condensate approximation, we define the bubble surface tension as
\begin{equation}
    \Sigma = \int_{0}^{\infty} dr \left[\left(\frac{d \sigma}{dr} \right)^2 - \frac{1}{2}\left(\frac{d \omega}{dr} \right)^2 - \frac{1}{2}\left(\frac{d \rho}{dr} \right)^2 - \frac{1}{2}\left(\frac{d \phi}{dr} \right)^2 \right].
\end{equation}

Figure \ref{fig:bubbles} illustrates the behavior of the surface tension in the spinodal region, which has been normalized by the corresponding critical chemical potential $\mu_{c}$ in each case. Since the critical bubble size diverges at the phase transition, the bubble computation is initiated at $\mu_{c}+2$MeV toward the end of the spinodal region, where $\Sigma = 0$. The solutions obtained within the domain wall approach, with $\mu_e = \mu_{e, \chi Broken}$, are denoted by the red squares, serving as a consistency check between the numerical methods employed. One can see a clear agreement between the results obtained from both approaches, validating their reliability. Higher values of $m_{0}$ correspond to narrower spinodal regions and smaller magnitudes of $\Sigma$.

\subsubsection{Mixed phases}

So far, we have described a first-order phase transition using the Maxwell construction. However, due to the relatively low values of surface tension observed in the previous section, there is a possibility that mixed phases, rather than homogeneous matter, could emerge. For $m_{0} =460$ MeV, the value of surface tension is $\Sigma = 0.32$ MeV/$\text{fm}^2$ as seen in Figure \ref{fig:surfacetension}. To explore this possibility, we adopt the Gibbs construction, relaxing the local charge neutrality condition to global conditions.

\begin{figure}
\begin{center}
\includegraphics[width=0.65\textwidth]{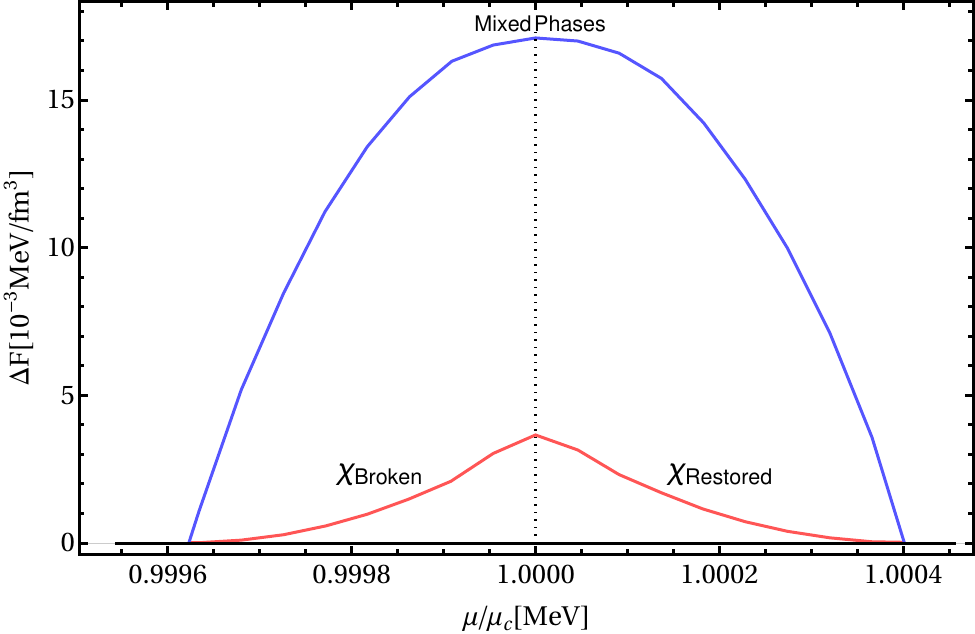}
\caption{Free energy densities of mixed phases (purple curve) and homogeneous phases (red curve) relative to the Gibbs constructed solutions (black horizontal line) for $\Sigma = 0.32$ MeV/$\text{fm}^2$. The dotted line marks the critical baryon chemical potential.}
\label{fig:mixedphases}
\end{center}
\end{figure}

In the Gibbs construction, we disregard the Coulomb and surface energies of the crystal structures and focus on calculating the energy for solutions with a nonspecified structure but with a chirally restored volume fraction $\chi$, with $\chi \in [0,1]$. The free energy density of the homogeneous solution (red curve) relative to the Gibbs constructed heterogeneous phases (black line) is illustrated in Fig. \ref{fig:mixedphases}. The dotted line marks the critical baryon chemical potential. We observe that the Gibbs constructed phase is only slightly favored when compared to Maxwell constructed solutions.

Including the Coulomb and surface corrections to the Gibbs solutions and searching for specific structures of various sizes such as cylinders, rods, and bubbles, that minimize the free energy density per unit volume, we find the mixed phase solution represented by the purple curve in Fig. \ref{fig:surfacetension}. Nevertheless, it becomes evident that despite the low values of surface tension, homogeneous phases are still energetically favored over mixed phases in this model. This outcome is attributed to the minimal energy gain in constructing heterogeneous phases, as demonstrated by the marginal difference in energy density between the Gibbs and Maxwell constructed phase transitions.

This finding should be compared to the previously encountered mixed phases found to be favored in neutron matter near the chiral phase transition \cite{Schmitt:2020tac}. The introduction of hyperons is expected to hinder the occurrence of mixed phases, as the additional baryonic hyperon degrees of freedom alleviate the energy cost of isospin asymmetry in charge-neutral homogeneous phases, lowering their energy cost in relation to the Gibbs solutions. The same effect occurs, for example, in hadron-quark mixed phases \cite{PhysRevC.52.2250}. In this study, we observe a different scenario, where the mixed phases consist of a chirally broken nucleonic phase and chirally restored parity-doubled neutrons (see the left panel of Fig. \ref{fig:nopds}). Hyperons only appear at significantly higher densities, so they cannot explain the disappearance of mixed phases. Apart from these differences, parity doubling and the presence of an invariant mass $m_{0}$ also act to prevent the formation of mixed phases.

\subsection{Neutron stars}
\subsubsection{Hyperonization and parity partners}

\begin{figure}
\begin{center}
\includegraphics[width=0.48\textwidth]{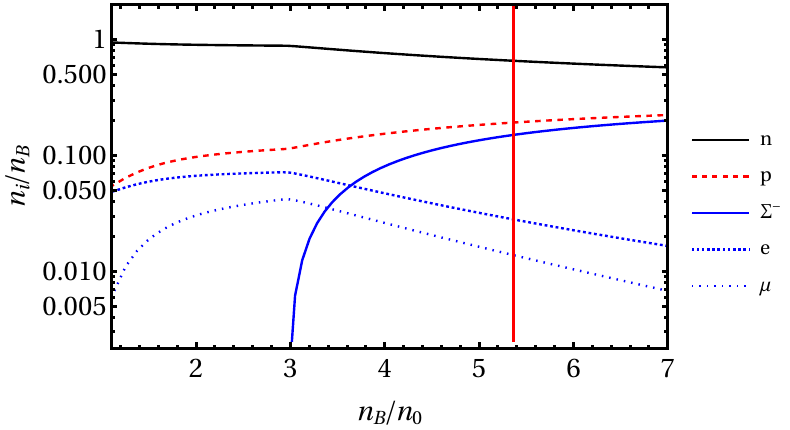}
\includegraphics[width=0.48\textwidth]{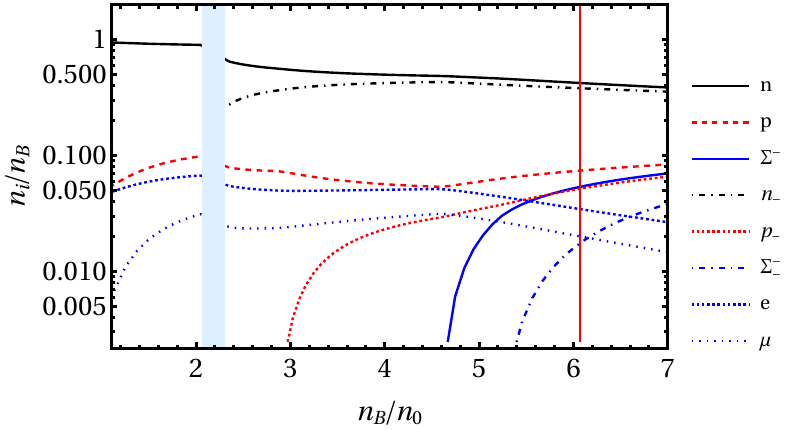}
\caption{Particle fractions for $M_{0}/M_{N} = 0.65$ and $m_{0} = 460$ MeV. On the left panel, parity partners are decoupled and matter consists solely of the baryon octet and leptons. On the right panel all parity partners are included. The blue band represents the discontinuity associated with the phase transition, while the red vertical lines indicate the maximum density reached within the interiors of the most massive stable stars built using the respective equations of state.}
\label{fig:nopds}
\end{center}
\end{figure}

In this section, we present our results regarding the impact of parity doubling on the composition of neutron stars and the stiffness of the equation of state. Figure \ref{fig:nopds} compares the resulting particle fraction with and without parity doubling for zero-temperature, beta-equilibrated matter using the same parameters of $M_{0}/M_{N} = 0.65$ and $m_{0} = 460$ MeV. On the left panel, we prevent the appearance of negative parity partners by decoupling the negative baryon states by assigning an extremely high mass to these states. As a result, their onset are artificially pushed to very high chemical potentials. On the right panel, we allow the onset of all parity partners by keeping their vacuum masses as described in Sec. \ref{subsec:parameterfixing}. We observe that the inclusion of negative parity states induces a first-order phase transition (marked by the blue band), with the N(1535) partner of the neutron emerging as the second most abundant baryonic state immediately after the phase transition. Additionally, the onset of strangeness, indicated by the presence of $\Sigma$ baryons, is shifted to significantly higher densities, transitioning from approximately $n_{B}/n_{0} = 3.02$ to $n_{B}/n_{0} = 4.68$. Moreover, the particle fraction of strangeness is reduced in relation to the amount of its value in when parity doubling is excluded, going from around $10 \%$ to around $7 \%$ at the maximum density achieved by the most massive stable stars built using the corresponding equations of state, which corresponds to the red vertical red line in the figure. We also observe that this line is pushed to higher densities when parity doubling is considered, which makes the cores of those stars more densely packed with baryons.

\begin{figure}
\begin{center}
\includegraphics[width=0.45\textwidth]{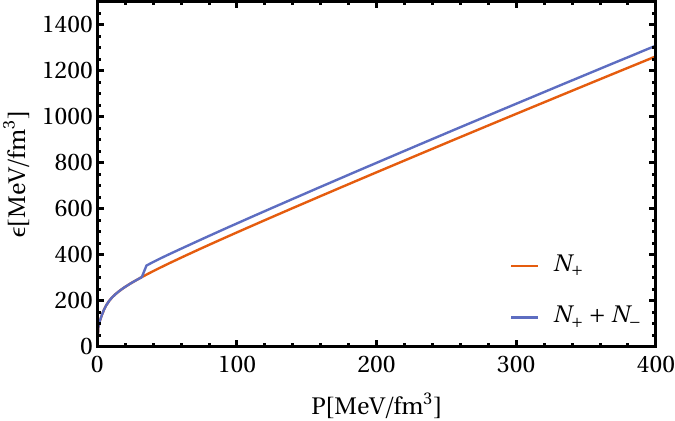}
\includegraphics[width=0.45\textwidth]{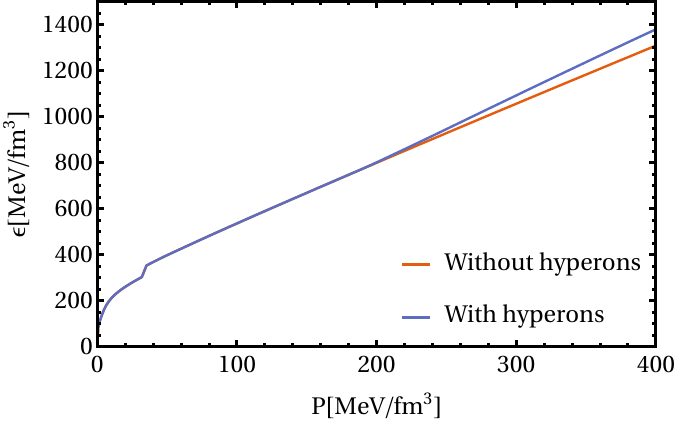}
\caption{Comparing parity doubling and hyperonization effects on the EoS. Left panel: EoS for nucleons and N(1535) states, revealing pressure reduction with parity doubling onset due to increased degrees of freedom post-transition. Parity doubling maintains consistent EoS slope through identical vector meson coupling. Right panel: Hyperonization-induced softening, stronger with rising strange baryon presence. Both positive and negative parity baryons were included. Weak hyperon-vector coupling eases repulsion, lowering pressure and speed of sound. All EoS were computed for the same parameters and couplings.}
\label{fig:eos}
\end{center}
\end{figure}

To distinguish between the impact of parity doubling and hyperonization on the equation of state (EoS), we compare these two scenarios in Figure \ref{fig:eos}. On the left panel, we present the EoS considering only nucleons and allow for the appearance of N(1535) states but no hyperons. Just after the phase transition, when the N(1535) states onsets, the parity doubled EoS has less pressure for the same energy density of the case without parity doublets, which is related to the fact that after the transition the system has more degrees of freedom to fill up as the chemical potential rises. 

A significant aspect of parity doubling within this model is the preservation of a relatively consistent slope in the EoS, similar to the scenario involving only nucleons. This uniformity arises due to the identical coupling with the vector meson sector for both parity-based baryonic states. Consequently, the contribution of N(1535) to the repulsive component of the nuclear force becomes comparable to that of the nucleon. Correspondingly, when considering a specific energy density, the speed of sound squared, denoted as $c_{s}^2= \frac{\partial P}{\partial \epsilon}$, remains indistinguishable between the two setups. On the right panel, we observe softening caused by the onset of hyperons with the inclusion of both positive and negative parity baryons. In this case the onset of strangeness is continuous and the softening becomes more pronounced as the fractional amount of strange baryons in the system increases for higher energy densities. This progressive softening stems from the fact that hyperons tend to couple weakly to the vector sector alleviating the repulsion and, consequently, the pressure. This results in a decrease in the value of $c_{s}^2$ as pressure increases, in contrast to the scenario where strangeness is absent. It is important to note that all equations of state were computed using the same parameters, which leads to the same couplings. Also notice that the softening resulting from hyperonization occurs relatively late. This delay is due to the fact that parity doubling, as explained previously, tends to push hyperonization to higher densities.

\begin{figure}
\begin{center}
\includegraphics[width=0.45\textwidth]{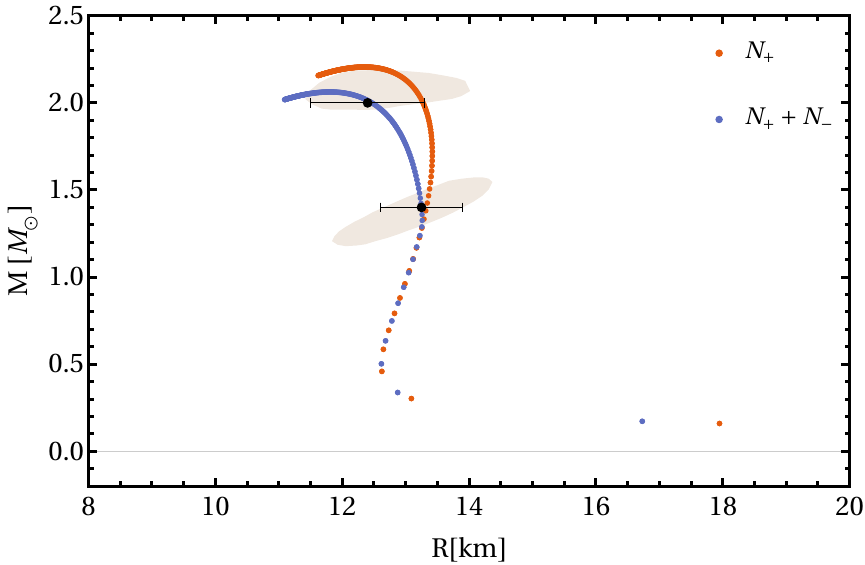}
\includegraphics[width=0.45\textwidth]{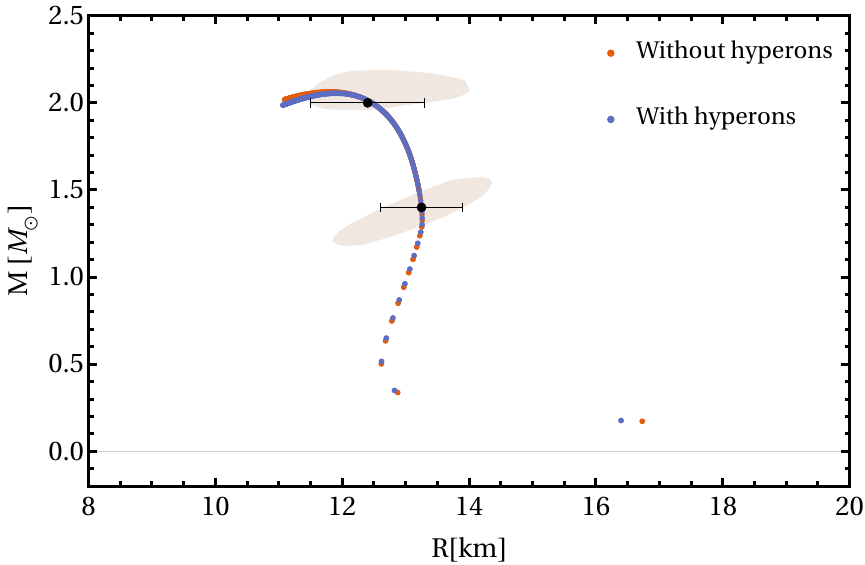}
\caption{Mass-radius diagrams from EoS in Figure \ref{fig:eos}. Gray ellipses: one-sigma confidence NICER radii measurements. Black bars: radii at 1.4 and 2.0 solar masses. Left: softening due to parity doubling, affecting larger stars; minimal impact on 1.4 solar mass stars. All curves stay NICER data compatible. Right: hyperonization across baryon octet, both parities. Hyperons at high densities impact $> 2.0$ solar mass stars, suggesting small hyperonization core region.}
\label{fig:mrs}
\end{center}
\end{figure}

The stellar configurations corresponding to the equations of state (EoS) shown in Fig. \ref{fig:eos} are depicted in Fig. \ref{fig:mrs}. 
The gray ellipses represent the results of combined one-sigma confidence analyses extracted from the studies \cite{Miller:2019cac, Riley:2019yda, Miller:2021qha, Riley:2021pdl}. The black bars indicate the radius values of $1.4$ and $2.0$ solar masses.

On the left panel, the EoS exhibits a noticeable softening effect due to parity doubling, particularly in the context of nucleons. This softening effect has a more pronounced impact on the more massive stars, whereas it has a minimal effect on the $1.4$ solar mass star, which has a relatively small chirally restored core. Despite this overall reduction in radii, none of the curves deviate to an extent that would render them incompatible with NICER data. In our analyses, this compatibility is represented by a mass-radius configuration that remains within the elliptical regions.

In contrast, the right panel illustrates the inclusion of hyperonization across the entire baryon octet, incorporating both parities. The inclusion of hyperons does not lead to a significant alteration in the maximum mass when parity doubling is present. Hyperons emerge only at high densities, thereby impacting solely the most massive stars with masses exceeding 2.0 solar masses. Consequently, in this scenario it is anticipated that only the most massive stars will have a small core region where hyperonization can occur.

\begin{figure}
\begin{center}
\includegraphics[width=0.65 \textwidth]{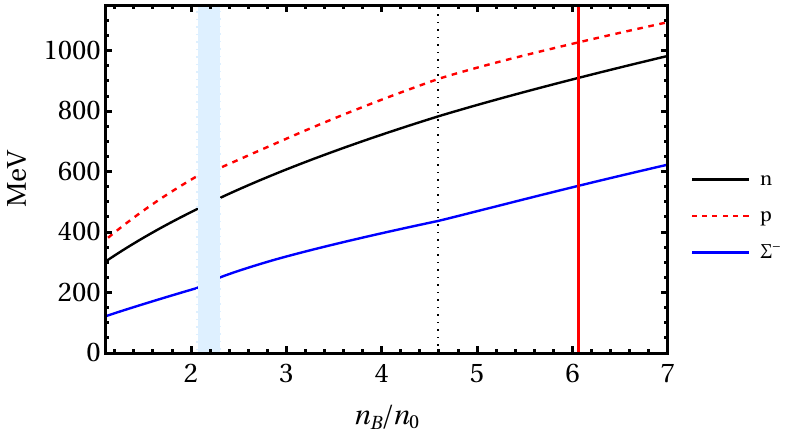}
\caption{Y-axis: $g_{i, \omega} \omega + g_{i, \rho} \rho + g_{i, \phi} + \mu_{i, \text{charge}}$
, with $i = n, p, \Sigma$.
A vertical red line indicates the maximum density achieved in the most massive stable star. Additionally, a vertical dotted line represents the onset of the $\Sigma^{-}$ baryon, the sole strange baryon onset in our model.
The plotted quantity serves to quantify the increased repulsion within the system as the number of the \textit{i}th species is elevated for a given density.}
\label{vecrep}
\end{center}
\end{figure}

In Fig. \ref{vecrep}, we compare the increased repulsion within the system for a given density. As in this work we assume that negative parity partners couple in the same way with the vector meson sector, these curves are identical for corresponding parity partners. The vertical red line indicates the maximum density achieved in the most massive stable star. Additionally, the vertical dotted line represents the onset of the $\Sigma^{-}$ baryon, the only hyperon that onsets in our model. For a single condensate, e.g. $\omega$ condensate, a simple comparison between $g_{i, \omega}$ suffices. However, given the presence of $\rho$ and $\phi$ condensates, the plotted quantity serves to quantify the increased repulsion within the system as the number of the \textit{i}th species is increased. This analysis provides insight into the dynamic interplay of the repulsive forces introduced by each species in the system. The figure makes it clear that the addition of $\Sigma^{-}$ baryons contributes significantly less to the overall repulsion of the system than the addition of nucleons (and their parity partners). Supporting the main conclusion, while both hyperonization and parity doubling soften the equation of state due to the extra Fermi levels available, hyperonization softens relatively more when it comes to repulsive interactions alone. As parity doubling tends to push hyperonization to occur at higher densities, its net effect is to stiffen the EoS when compared to what is expected from hyperonization alone, without parity doubling.

\subsubsection{Metastable cores}

\begin{figure}
\begin{center}
\includegraphics[width=0.45\textwidth]{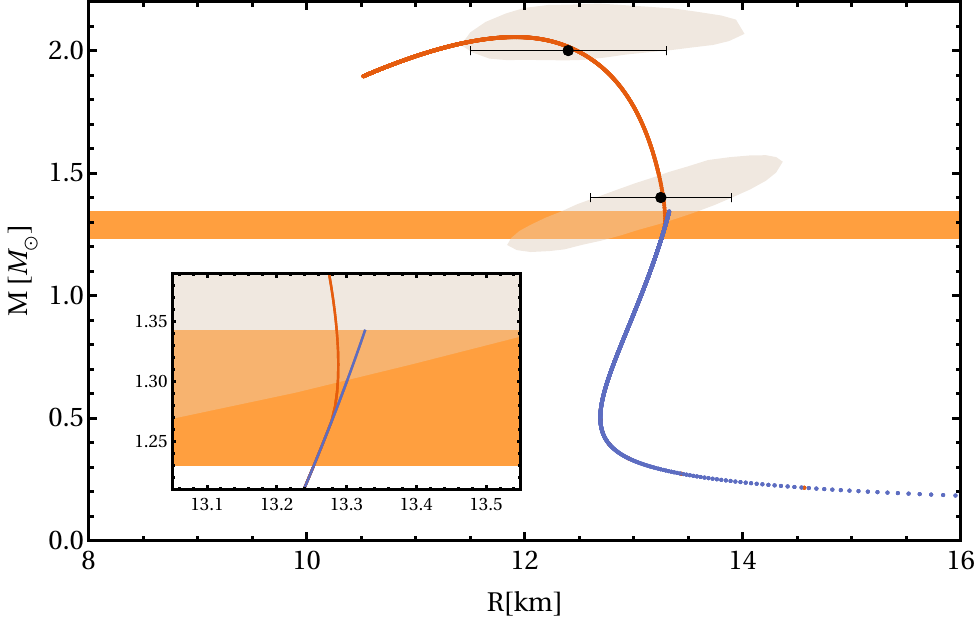}
\includegraphics[width=0.45\textwidth]{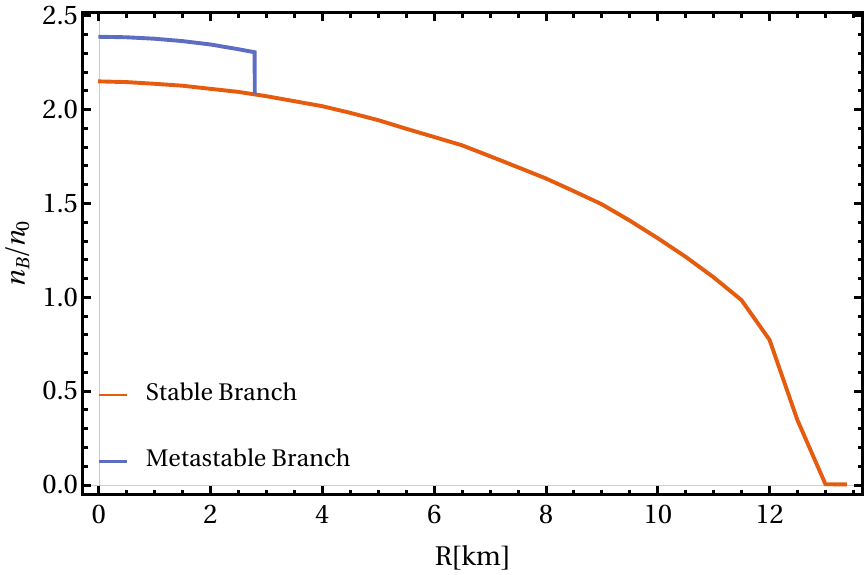}
\caption{Left panel: mass-radius relations for the stable and metastable branches, distinguished by the orange and blue curves, respectively. The shaded orange band represents the mass range within which both configurations coexist. The gray ellipses correspond to constraints from NICER with black bars marking radii of 1.4 and 2.0 solar mass stars. Right panel: baryon density vs. star radius for stars with an identical baryonic mass of 1.3 solar masses. These stars are built using solutions from both branches of the equation of state, resulting in distinct density-radius profiles at the core.}
\label{fig:metacore}
\end{center}
\end{figure}

The advantage of having a single description for both the chirally restored and chirally broken phase  is that phenomena associated with metastability and their timescales can be investigated in a unified framework. We see from Figs. \ref{fig:tunability} and \ref{fig:bubbles} that, in the cases where the chiral restoration occurs via a first-order phase transition, the spinodal region can extend through a considerable range of baryon chemical potential. In Fig. \ref{fig:metacore} we show the resulting family of stars obtained by solving the TOV equations for the EoS corresponding to the black parameter set of Table \ref{table:para}. If we follow the thermodynamically stable branches through the phase transition, we branch off of the neutron star curve by following the chirally restored branch. We obtain hybrid stars, shown by the orange curves in Fig. \ref{fig:metacore}, i.e., stars with a chirally broken mantle and a chirally restored core. If instead we follow the thermodynamically metastable solution after the phase transition we obtain the purple curve, where all the star is formed by matter in its chirally broken phase, but in the core matter is in a metastable phase (see Fig. \ref{fig:metacore}). The orange band highlights the mass range at which both solutions are possible and we can encounter this mass degenerate star configurations. These bands begin when the critical central pressure for phase transition is reached at the center of the star and end where the metastable branch reaches the spinodal point and no metastable solution exists for greater central pressure.

The gravitational and baryonic masses are defined, as usual, by \cite{glendenning2012compact} 
\begin{equation}
    M(r) = 4 \pi \int_{0}^{r} \epsilon(r') r'^2 dr' \,,
\end{equation}
so that the gravitational mass is defined as $M_{G} \equiv M(R)$ and
\begin{equation}
M_{A} \equiv \sum_{i,p=\pm} A_{i,p} m_{i,p}^{vac}\, ,    
\end{equation}
with the baryonic number for each species $A_{i,p}$ given by
\begin{equation}
    A_{i,p} \equiv 4 \pi \int_{0}^{R} \left(1- \frac{2 M(r)}{r} \right)^{-1/2}r^2 \rho_{i,p}(r)dr \, . 
\end{equation}

\begin{figure}
\begin{center}
\includegraphics[width=0.45\textwidth]{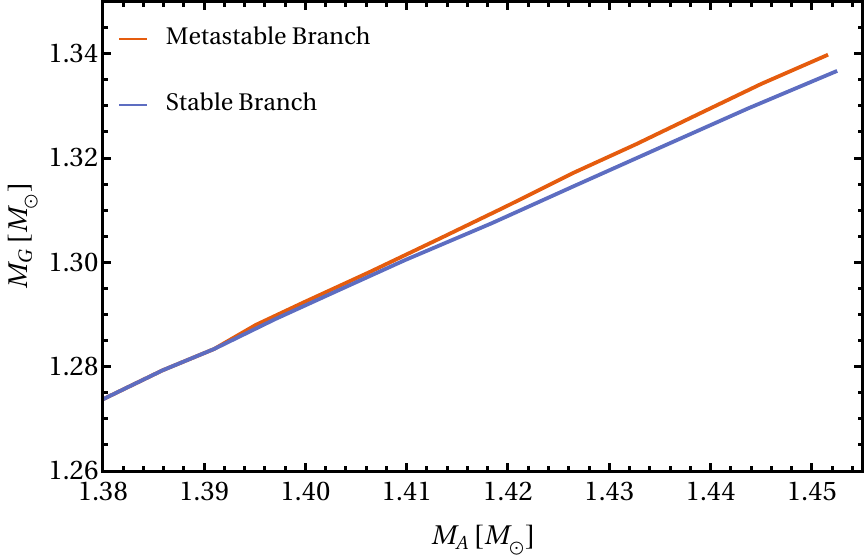}
\includegraphics[width=0.45\textwidth]{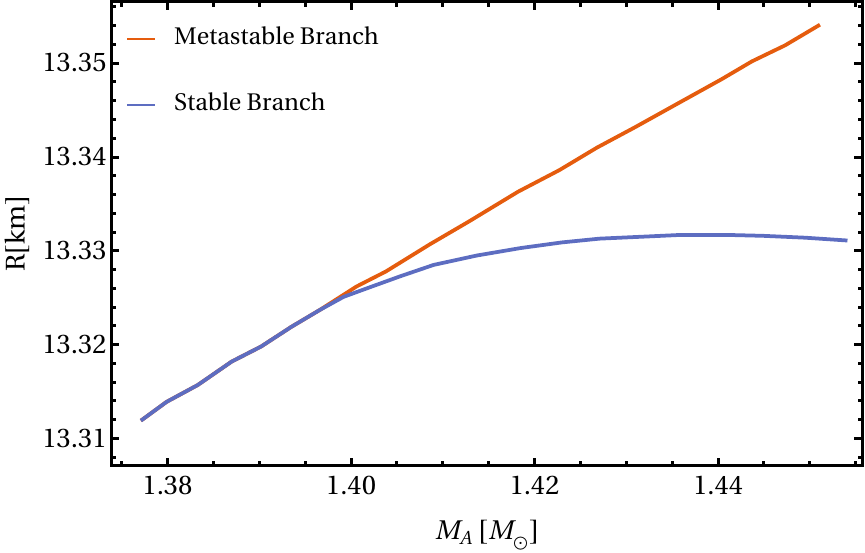}
\caption{Gravitation mass (left panel) and radius (right panel) for star configurations with the same baryonic mass. Both branches correspond to the metastable and stable EoS obtained for parameters $M_{0}/M_{N} = 0.65$ and $m_{0} = 460$ MeV.}
\label{fig:metastar}
\end{center}
\end{figure}

Since no known physical process violates baryon number conservation in compact stars, $M_{A}$ should be conserved in isolated stars. Thus, the difference between baryonic mass and gravitational mass can be used as a measurement of the binding energy of the star $B = M_{A}-M_{G}$. We have checked that for the same baryonic mass the change in gravitational mass associated with the core phase conversion is $\lesssim 0.02 \%$ as shown on the left panel of Fig. \ref{fig:metastar} for the metastable stars and for the completely stable stars corresponding to the orange band of Fig. \ref{fig:metacore}. This energy difference, associated with the change in binding energy, could in principle drive a late heating and late emissions from the star, depending on the timescales for the phase transition. 
On the right panel of Fig. \ref{fig:metastar}, the difference in radii between metastable and stable stars as a function of baryonic mass is shown. We see that  metastable stars are only slightly larger than the correspondingly (same gravitational mass) stable stars, and the order of magnitude of the radius difference is $\Delta R \lesssim 20$m.

Given the relatively small values of surface tension, as seen in Sec. \ref{sec:surfacetension}, it is not unreasonable to anticipate that these metastable stars would have a limited lifespan. Nevertheless, the metastable phases can be relevant in the early development of protoneutron stars. Additionally, the relatively short timescales involved indicate that, in scenarios characterized by dynamical processes, where localized areas can undergo destabilization due to fluctuations in pressure or lepton number, the transition from a state of chirally restored matter to a state of chirally broken matter (or vice-versa) would be constrained primarily by the timescales of weak decay that drives beta equilibration.

To illustrate this point, consider the case of a neutron star undergoing vibrations. These oscillations lead to variations in pressure, with the response differing according to the timescales of phase conversion. Another instance in which such dynamical configurations can manifest is in the aftermath of compact star mergers and in core-collapse supernovae. In these scenarios, regions of lepton-rich matter can stabilize areas of chirally broken symmetry. As the system cools and progresses toward lower leptonic densities and temperatures, these regions can destabilize, and transition to the chirally restored phase in a timeframe determined by weak interactions.

\section{Summary and outlook}
\label{sec:summary}

We have investigated the impact of chiral symmetry restoration through parity doubling in beta-equilibrated cold matter. To achieve this, we built a chiral Lagrangian that incorporates the baryon octet and their corresponding negative parity partners. A crucial aspect of this approach is the enabling of a chiral-invariant mass $m_{0}$, which is possible because of the specific way the opposite parity baryon fields transform. Consequently, the chirally symmetric phase is not solely composed of massless baryons. Our model was calibrated to reproduce nuclear matter properties at saturation density. By exploring the parameter space, we identified certain parameter sets that satisfy the mass and radius constraints imposed by recent astronomical observations without the need to include a quark matter phase.

The early onset of the N(1535) resonance, which occurs immediately after the phase transition, inhibits strangeness formation and leads to hyperonization at higher densities compared to standard chiral restoration schemes for baryonic matter, which constitutes another solution to the ``hyperon puzzle'' problem.

We found a relatively subtle first-order transition into the chirally restored phase, with position of the critical density and energy density discontinuity mainly dictated by the chiral-invariant mass $m_{0}$. The corresponding surface tension is one order of magnitude smaller than prior results using similar baryon-meson approaches. It exhibits a strong dependence on the value of $m_{0}$, given its direct correlation to the alteration in effective baryonic mass across the phase transition. We employed the Maxwell construction, which remains applicable even for low surface tension values. Its validity stems from the energy associated with forming mixed phases, predominantly governed by the Coulomb contribution arising from the global charge neutrality condition. This condition favors the existence of a sharp interface between phases at the critical chemical baryon potential.

Investigating the impacts of parity doubling and hyperonization on the EoS led us to conclude that parity doubling significantly influences stellar radii of the most massive stars, generally causing their reduction. However, this effect is not big enough to be resolved by NICER measurements since the change in radii is smaller then the one-sigma ellipses.

By utilizing both the metastable and stable branches of the equation of state, we compared stars containing a metastable core to those composed exclusively of the most stable homogeneous solutions. Despite the apparently short lifespans for these metastable cores at finite temperatures due to their low surface tension values, the phase conversion and subsequent contraction of stars could potentially influence dynamical phenomena in neutron stars with masses around 1.3 solar masses or greater. 
In dynamical scenarios, where local regions could destabilize due to pressure or lepton fluctuations, the transition between chirally restored and chirally broken matter depends on weak decay timescales. For example, neutron star vibrations induce pressure changes with responses based on phase conversion timescales. Similarly, after compact star mergers and in core-collapse supernovae, lepton-rich zones stabilize chirally broken symmetry areas. As systems cool and densities drop, these stable regions may destabilize and shift to the restored phase in a time frame controlled by weak interactions. A detailed calculation of these timescales will be left for a future investigation.

One generalization of this model concerns the physical interpretation of the chiral-invariant mass $m_{0}$. Possible dynamical ways of generating this term come from interactions with tetraquark condensates and the gluon condensate \cite{Heinz:2013hza,PhysRevD.82.014004}. Additionally, the inclusion of the $\zeta$ condensate as an independent parameter, or the consideration of different monotonic functions $\zeta(\sigma) = f(\sigma)$ instead of simple identification, is desirable.

Another relevant issue involves the inclusion of the vacuum contribution, commonly referred to as the ``Dirac sea." This contribution is usually disregarded within relativistic mean field models applied to neutron stars and nucler physics, since vacuum effects were shown to have minimal impact on the equation of state \cite{GLENDENNING1989521}. However, in a variety of models involving chiral symmetry restoration, it is known that a first-order chiral transition can be smoothed to a crossover when vacuum terms are included, bringing qualitative modifications \cite{Fraga:2008qn,Mizher:2010zb,Endrodi:2013cs,Haber:2014ula}, 
for a recent work on chiral density waves where the
nucleonic vacuum contributions are included in a similar model see \cite{Pitsinigkos:2023xee}.

Our parity doubling framework can be extended to nonzero temperatures and scenarios beyond beta equilibrium. A crucial aspect in this context is the examination of the direct URCA process, which involves rapid neutrino emission in neutron stars. Neutrons transform into protons, releasing an electron and an electron antineutrino, while protons capture electrons, becoming neutrons and emitting electron neutrinos. This cooling mechanism carries away thermal energy via diffusion and radiation of neutrinos and antineutrinos. Its occurrence depends on specific conditions, with a critical proton fraction denoted as $Y_{p}^{\text{DU}}$. Notably, parity doubling modifies this critical value leading to an overall reduction of this critical proton threshold, as suggested in Ref. \cite{PhysRevD.98.103021}.

Expanding this model to finite temperatures also offers opportunities to explore the chiral phase transition in core-collapse supernovae, the evolution of protoneutron stars and neutron star mergers. In core-collapse supernovae, gravitational pressure cause the collapse of massive stars. The interplay between chiral symmetry restoration and exotic hadronic state formation significantly influences the explosion mechanism and subsequent nucleosynthesis. Neutron star mergers involve extreme conditions like high temperatures and rapid density changes. Understanding the state of nuclear matter formed in such processes, its phase transitions and their impact on gravitational wave emission and heavy element production through r-process nucleosynthesis is pivotal. This is especially timely given the ongoing operations of LIGO/Virgo/Kagra in the current O4 run \cite{Cahillane:2022pqm}, promising increased statistics and precision for gravitational wave signals from neutron star mergers.

\section{Acknowledgments}
We thank M. Alford, A. Motornenko, S. Pitsinigkos, A. Schmitt and J. Steinheimer for helpful comments and discussions. E.S.F. is partially supported by CAPES
(Finance Code 001), CNPq, FAPERJ, and INCT-FNA (Process No. 464898/2014-5). R.M. thanks CNPq (Process No. 200231/2022-7) for financial support. R.M. and J.S.B. acknowledge support by the Deutsche Forschungsgemeinschaft (DFG, German Research Foundation) 
through the CRC-TR 211 `Strong-interaction matter under extreme conditions'– project number 315477589 – TRR 211.
\section{Appendix A}\label{appendix:a}

In this appendix, we first explicitly present the principal aspects of the diagonalization of the SU(3) model calculation, which leads to the masses as given in Eqs. (\ref{massnucleon}), (\ref{masslambda}), (\ref{masssigma}), (\ref{masscascade}). This diagonalization occurs in the Scalar meson-baryon sector. Finally, we shift our focus to the Vector meson-baryon sector and establish the constraints between vector couplings imposed by SU(3) symmetry.

\subsection*{1. Scalar mesons - baryon sector}\label{appendix:aa}

In the nonlinear realization of SU(3) chiral symmetry, the various interaction terms of baryons with mesons 
have the same structure, except for the difference in Lorentz space \cite{PhysRevC.57.2576,Papazoglou:1998vr}. For a general meson field $W$ they read
\begin{equation}\label{baryonmesonint}
\mathcal{L}_{BW}=- \sqrt{2}g_{8}^{W}(\alpha_{W}\left[\bar{\Psi} \mathcal{O} \Psi \right]_{F} + (1-\alpha_{W})\left[\bar{\Psi} \mathcal{O} \Psi W\right]_{D}) -g_{1}^{w}\frac{1}{\sqrt{3}} \Tr{(\bar{\Psi}\mathcal{O} \Psi)}\Tr{W} \,,
\end{equation}
where $\left[\bar{\Psi} \mathcal{O} \Psi \right]_{F} := \Tr{(\bar{\Psi}\mathcal{O}W \Psi -\bar{\Psi}\mathcal{O}\Psi W)}$ and $\left[\bar{B} \mathcal{O} \Psi \right]_{D} := \Tr{(\bar{\Psi}\mathcal{O}W \Psi +\bar{\Psi}\mathcal{O} \Psi W)}-\frac{2}{3}\Tr{W}$.
By a redefinition the parameters $g_{8}^{W}$, $g_{1}^{W}$ and $\alpha_{W}$ Eq. (\ref{baryonmesonint}) is rewritten as
\begin{equation}\label{baryonintrewrite}
\mathcal{L}_{BW}= D_{W} \Tr(\bar{\Psi}\mathcal{O}\{W,\Psi \}) + F_{W} \Tr(\bar{\Psi}\mathcal{O}\left[W,\Psi\right]) + S_{W} \Tr{(\bar{\Psi}\mathcal{O} \Psi)}\Tr{W} \,.
\end{equation}

For the interaction between baryons and scalar (and pseudoscalar) mesons, one takes $W=X$ and $\mathcal{O}=1$. The baryon degrees of freedom are parameterized as usual
\begin{equation}\label{baryonoctet}
\Psi =\left(\begin{array}{ccc} \displaystyle{\frac{\Sigma^0}{\sqrt{2}} + \frac{\Lambda}{\sqrt{6}}} & \Sigma^+ & p \\ \Sigma^- &
\displaystyle{-\frac{\Sigma^0}{\sqrt{2}} + \frac{\Lambda}{\sqrt{6}}}  & n  \\ \Xi^- & \Xi^0 & \displaystyle{-\sqrt{\frac{2}{3}}\Lambda} 
\end{array}\right) \, .
\end{equation}

The scalar and pseudoscalar meson nonets are encoded in the field $X  =  \Sigma+i\Pi = T_a(\sigma_a+i\pi_a)$, where $T_a=\lambda_a/2$ for $a=0,\ldots, 8$, with  the Gell-Mann matrices $\lambda_a$ for $a=1,\ldots,8$ and $\lambda_0 = \sqrt{2/3}\,{\bf 1}$. This is usually reparametrized as 
\begin{subequations}
\begin{eqnarray}
\Sigma =T_a\sigma_a &=& \frac{1}{\sqrt{2}}\left(\begin{array}{ccc} \displaystyle{\frac{a_0^0}{\sqrt{2}} + \frac{\sigma_8}{\sqrt{6}}+\frac{\sigma_0}{\sqrt{3}}} & a_0^+ & \kappa^+ \\ a_0^- &
\displaystyle{-\frac{a_0^0}{\sqrt{2}} + \frac{\sigma_8}{\sqrt{6}}+\frac{\sigma_0}{\sqrt{3}}}  & \kappa^0  \\ \kappa^- & \bar{\kappa}^0 & \displaystyle{-\sqrt{\frac{2}{3}}\sigma_8+\frac{\sigma_0}{\sqrt{3}}} 
\end{array}\right) \, , \\[2ex]
\Pi=T_a\pi_a &=& \frac{1}{\sqrt{2}}\left(\begin{array}{ccc} \displaystyle{\frac{\pi^0}{\sqrt{2}} + \frac{\pi_8}{\sqrt{6}}+\frac{\pi_0}{\sqrt{3}}} & \pi^+ & K^+ \\ \pi^- &
\displaystyle{-\frac{\pi^0}{\sqrt{2}} + \frac{\pi_8}{\sqrt{6}}+\frac{\pi_0}{\sqrt{3}}}  & K^0  \\ K^- & \bar{K}^0 & \displaystyle{-\sqrt{\frac{2}{3}}\pi_8+\frac{\pi_0}{\sqrt{3}}} 
\end{array}\right)   \, .
\end{eqnarray}
\end{subequations} 
In the scalar sector, one can trade $\sigma_0$ and $\sigma_8$ for 
nonstrange and strange scalar fields by the transformation 
\begin{equation} \label{trafo}
\left(\begin{array}{c} \sigma \\ \zeta\end{array}\right) = \frac{1}{\sqrt{3}}\left(\begin{array}{cc} \sqrt{2} & 1 \\ 1 & -\sqrt{2} \end{array}\right)\left(\begin{array}{c} \sigma_0 \\ \sigma_8\end{array}\right) \, . 
\end{equation}
At mean field level ($\langle \pi_{a} \rangle = 0$, $ \langle\sigma_{0}\rangle =\sigma_{0}$ and $ \langle\sigma_{8}\rangle =\sigma_{8}$), the Lagrangian of Eq. (\ref{intsym}) in terms of the transformed fields of Eq. (\ref{transformedfield}) leads to the masses of Eqs.(\ref{massnucleon}),(\ref{masslambda}), (\ref{masssigma}), (\ref{masscascade}) for $\delta_{ij}$ defined implicitly as
\begin{subequations}
\begin{align}\label{deltaij}
\sinh(\delta_{11})&=\sinh(\delta_{12})=\sinh(\delta_{21})= \sinh(\delta_{22}) =\frac{1}{m_{0}} \left[\left(D_{s}^{(1)}+S_{s}^{(1)}\right) \sigma +\frac{\sqrt{2}}{2}\left(S_{s}^{(1)}\right)\zeta\right] \, , \\
\sinh(\delta_{13})&=\sinh(\delta_{23})= \frac{1}{m_{0}} \left[\left(\displaystyle{\frac{D_{s}^{(1)}+F_{s}^{(1)}}{2} + S_{s}^{(1)}}\right)\sigma + \frac{\sqrt{2}}{2}\left(D_{s}^{(1)}-F_{s}^{(1)} +S_{s}^{(1)}\right)\zeta \right]\, , \\
\sinh(\delta_{31})&=\sinh(\delta_{31})= \frac{1}{m_{0}} \left[\left(\displaystyle{\frac{D_{s}^{(1)}-F_{s}^{(1)}}{2} + S_{s}^{(1)}}\right)\sigma + \frac{\sqrt{2}}{2}\left(D_{s}^{(1)}+F_{s}^{(1)} +S_{s}^{(1)}\right)\zeta \right] \, , \\
\sinh(\delta_{33})&= \frac{1}{m_{0}} \left[\left( S_{s}^{(1)}\right)\sigma + \frac{\sqrt{2}}{2}\left(2 D_{s}^{(1)}+S_{s}^{(1)}\right)\zeta \right] \, .
\end{align}
\end{subequations}

\subsection*{2. Vector mesons - baryon sector}\label{appendix:ab}

We assume that vector meson couple similarly to both baryon parity state. The interaction term has the same structure of Eq. (\ref{baryonintrewrite}), but with $W=V_{\mu}$ and $\mathcal{O}=\gamma^{\mu}$, with
\begin{equation}
    \mathcal{L}_{BV} = D_{v} \Tr \left(\bar{\Psi} \gamma^{0} \{V_{0},\Psi \}  \right) + F_{v} \Tr \left(\bar{\Psi} \gamma^{0} \left[V_{0},\Psi \right]  \right) + S_{v} \Tr \left( V_{0} \right) \Tr \left(\bar{\Psi} \gamma^{0} \Psi \right)\, .
\end{equation}
Similarly to the baryon-scalar sector, the 11 couplings  with the vector condensates $\{g_{N v}, g_{\Lambda v},g_{\Sigma v},g_{\Xi v}\}$ ($v=\omega$,$\rho$,$\phi$) are linear combinations of the 3 independent coefficients $D_v,F_v,S_v$. Choosing three couplings, say the 3 nucleonic couplings $g_{N\omega}$, $g_{N\phi}$, $g_{N\rho}$, one can  express the remaining 8 hyperonic couplings as 
\begin{align}\label{vectorcoupling}
g_{\Sigma\omega} &= \frac{g_{N\omega}+\sqrt{2}g_{N\phi}+g_{N\rho}}{2}
\, , \qquad g_{\Lambda\omega} = \frac{5g_{N\omega}+\sqrt{2}g_{N\phi}-3g_{N\rho}}{6} \, , \qquad g_{\Xi\omega} = \frac{g_{N\omega}+\sqrt{2}g_{N\phi}-g_{N\rho}}{2} \, ,\\  \nonumber
g_{\Sigma\rho} &= \frac{g_{N\omega}-\sqrt{2}g_{N\phi}+g_{N\rho}}{2} \, , \qquad g_{\Lambda\rho} = 0 \, , \qquad g_{\Xi\rho} = \frac{g_{N\omega}-\sqrt{2}g_{N\phi}-g_{N\rho}}{2} \, ,\\  \nonumber
g_{\Sigma\phi} &= \frac{g_{N\omega}-g_{N\rho}}{\sqrt{2}} \, , \qquad g_{\Lambda\phi} = \frac{\sqrt{2}g_{N\omega}+4g_{N\phi}+3\sqrt{2}g_{N\rho}}{6} \, , \qquad g_{\Xi\phi} = \frac{g_{N\omega}+g_{N\rho}}{\sqrt{2}} \, .
\end{align}
%

\section{Appendix B}\label{appendix:b}

In this appendix, we provide an explanation of the parameter-fixing procedure used in this model. Most of the calculations are derived from Appendix A of Ref. \cite{Schmitt:2020tac}. However, it is still valuable to demonstrate how those ideas are applied to the current model. Aside from the similarities, certain differences also become apparent.
\newline

At saturation density (where there are no hyperons), we have $n_B=n_n+n_p$, $n_I=n_n-n_p$, and we shall need $\mu_B=(\mu_n+\mu_p)/2$, $\mu_I=(\mu_n-\mu_p)/2$ (which is true in general). The stationarity equations (\ref{stats}) are
\begin{subequations}
\begin{eqnarray}
0&=&\frac{\partial U}{\partial \sigma} + g_{N\sigma}(n_{sn}+n_{sp}) \, , \label{statsigmasat}\\[2ex]
0&=&\omega[m_\omega^2+d(\omega^2+3\rho^2)]-g_{N\omega}n_B\, , \label{statomegasat}\\[2ex]
0&=&\phi m_\phi^2-g_{N\phi}n_B\, , \label{statphisat}\\[2ex]
0&=&\rho[m_\rho^2+d(\rho^2+3\omega^2)]-g_{N\rho}n_I \, . \label{statrhosat}
\end{eqnarray}
\end{subequations}

For isospin-symmetric matter, we have $n_I=0$ and thus also $\rho=0$. In this case, we write the solutions for $\omega$ and $\phi$ as
\begin{equation} \label{omegaf}
\omega_0= \frac{g_{N\omega}n_0}{m_\omega^2}f(x_0)\, , 
\end{equation}
with
\begin{equation}
f(x) = \frac{3}{2x}\frac{1-(\sqrt{1+x^2}-x)^{2/3}}{(\sqrt{1+x^2}-x)^{1/3}} \, , \qquad x_0\equiv \frac{3\sqrt{3d}\,g_{N\omega}n_0}{2m_\omega^3} \, , 
\end{equation}
and
\begin{equation}
    \phi_{0}= \frac{g_{N\phi}n_0}{m_\phi^2}\, .
\end{equation}
With $\lim_{x\to 0} f(x)=1$ we recover the case without quartic vector meson interactions, $d=0$. The pressure of isospin-symmetric nuclear matter at saturation equals the vacuum pressure $P_{\rm vac}=0$ and can be written as
\begin{equation} \label{Psat}
0= P_{\rm sat} = \frac{m_\omega^2}{2}\omega_0^2+\frac{d}{4}\omega_0^4+\frac{m_\phi^2}{2}\phi_0^2-U(\sigma)+\frac{1}{4\pi^2}\left[\left(\frac{2}{3}k_F^3-M_0^2k_F\right)\mu_B^*+M_0^4\ln\frac{k_F+\mu_B^*}{M_0}\right]
\, ,
\end{equation}
where $\mu_B^* =\mu_n^*=\mu_p^* = \sqrt{k_F^2+M_0^2}$, and the Fermi momentum is given in terms of the saturation density via $n_0=2k_F^3/(3\pi^2)$, while the stationarity equation (\ref{statsigmasat}) at saturation for symmetric nuclear matter is
\begin{equation} \label{sigsat}
0=  \frac{\partial U}{\partial\sigma} + \frac{g_{N\sigma}M_0}{\pi^2}\left(k_F\mu_B^*-M_0^2\ln\frac{k_F+\mu_B^*}{M_0}\right) \, .
\end{equation} 
The relation $\mu_B^*=\mu_0-g_{N\omega}\omega_0-g_{N\phi}\phi_0$ together with Eq.\ (\ref{omegaf}) can be used to write $g_{N\omega}$ in terms of saturation properties and the coupling constant $d$,
\begin{equation} \label{gNw}
g_{N\omega}^2 = -\frac{g_{N\phi}^2 m_{\omega}^2}{2 m_{\phi}^2}+\frac{m_\omega^2}{2n_0}(\mu_0-\mu_B^*)+ \frac{1}{2 n_{0}}\sqrt{\frac{-4d g_{N \phi}^2 n_{0}^2+m_{\phi}^2\left(m_{\omega}^4 + 4 d n_{0} (\mu_{0}-\mu_{B}^{*})\right)\left(g_{N\phi}^2 n_{0}-m_{\phi}^2(\mu_{0}-\mu_{B}^{*})\right)^2}{m_\phi^6}} \, ,
\end{equation}
where $\mu_0=922.7\, {\rm MeV}$ is the value of $\mu_B$ at saturation. 

The incompressibility is defined as
\begin{equation}
K = 9n_B\frac{\partial\mu_B}{\partial n_B} \, ,
\end{equation}
where we assume isospin symmetry, $n_I=0$. We compute the derivative from 
\begin{equation}
\mu_B = g_{N\omega}\omega+g_{N\phi}\phi+\frac{\mu_n^*+\mu_p^*}{2} \, .
\end{equation}

For the first term, we insert Eq.\ (\ref{omegaf}) and take the derivative with respect to $n_B$, while for the second term we can simply follow appendix A of Ref.\ \cite{Schmitt:2020tac}. This yields
\begin{eqnarray}\label{incomp}
K &=& \frac{6k_F^3}{\pi^2}\left(\frac{g_{N\omega}^2}{m_\omega^2}[f(x)+xf'(x)]+\frac{g_{N\phi}^2}{m_\phi^2}\right)
+\frac{3k_F^2}{\mu_B^*}+\frac{9n_BM}{\mu_B^*}\frac{\partial M}{\partial n_B} \\ \nonumber
&=&\frac{6k_F^3}{\pi^2}\left(\frac{g_{N\omega}^2}{m_\omega^2}[f(x)+xf'(x)]+\frac{g_{N\phi}^2}{m_\phi^2}\right)
+\frac{3k_F^2}{\mu_B^*}-\frac{6k_F^3}{\pi^2}\left(\frac{M}{\mu_B^*}\right)^2
\left(\frac{\partial^2U}{\partial M^2}+\frac{2}{\pi^2}\int_0^{k_F} dk\,\frac{k^4}{\epsilon_k^3}\right)^{-1} \, .
\end{eqnarray}
Again, we can check that we recover the $d=0$ result with $f'(x)\propto x$ for small $x$. 

The symmetry energy is defined as
\begin{equation}
S =  \frac{n_B}{2} \frac{\partial\mu_I}{\partial n_I} \, ,
\end{equation}
where the derivative is taken at fixed $n_B$ and evaluated at $n_I=0$.
This derivative is computed from
\begin{equation}\label{muIsat}
\mu_I = g_{N\rho}\rho +\frac{\mu_n^*-\mu_p^*}{2} \, .
\end{equation}
We find from Eq.\ (\ref{statrhosat})
\begin{equation}
\left.\frac{\partial\rho}{\partial n_I}\right|_{\rho=\phi=0} = \frac{g_{N\rho}}{m_\rho^2+3d\omega^2} \, .
\end{equation}
For the derivative of the second term in Eq.\ (\ref{muIsat}) we can again simply follow appendix A of Ref.\ \cite{Schmitt:2020tac}. Consequently, 
\begin{equation}\label{Esym}
S = \frac{k_F^3}{3\pi^2}\frac{g_{N\rho}^2}{m_\rho^2+3d\omega^2}+\frac{k_F^2}{6\mu_B^*} \, . 
\end{equation}

With Eq.\ (\ref{omegaf}) we can use this result to write $g_{N\rho}$  as
\begin{equation} \label{gNrho}
g_{N\rho}^2 = \frac{3\pi^2m_\rho^2}{k_F^3}\left(S-\frac{k_F^2}{6\mu_B^*}\right)\left(1+\frac{3d\omega_0^2}{m_\rho^2}\right) \, .
\end{equation}
The slope parameter of the symmetry energy is defined as 
\begin{equation}
L = 3n_B\frac{\partial S}{\partial n_B} \, ,
\end{equation}
where the derivative is taken at fixed $n_I=0$, i.e., we can simply take the derivative of Eq.\ (\ref{Esym}) with respect to $n_B$. We compute
\begin{eqnarray}
\frac{\partial S}{\partial n_B} = \frac{1}{2}\frac{g_{N\rho}^2}{m_\rho^2+3d\omega^2}-\frac{3n_B g_{N\rho}^2d\omega}{(m_\rho^2+3d\omega^2)^2}\frac{\partial\omega}{\partial n_B}+\frac{\pi^2}{6k_F\mu_B^*}-\frac{\pi^2k_F}{12\mu_B^{*3}}-\frac{k_F^2M}{6\mu_B^{*3}}\frac{\partial M}{\partial n_B} \, .
\end{eqnarray}
We now use
\begin{equation}
\left.\frac{\partial\omega}{\partial n_B}\right|_{\rho=0} = \frac{g_{N\omega}}{m_\omega^2+3d\omega^2} \, ,
\end{equation}
which follows from Eq.\ (\ref{statomegasat}), and we express $\frac{\partial M}{\partial n_B}$ in terms of $K$ with the help of the first line of Eq.\ (\ref{incomp}). This yields
at saturation
\begin{equation}\label{L}
L = \frac{3g_{N\rho}^2n_0}{2(m_\rho^2+3d\omega_0^2)}\left[1-\frac{6d\,n_0g_{N\omega}\omega_0}{(m_\rho^2+3d\omega_0^2)(m_\omega^2+3d\omega_0^2)}\right]+\frac{k_F^2}{3\mu_B^*}\left(1-\frac{K}{6\mu_B^*}\right)+\frac{n_0k_F^2}{2\mu_B^{*2}}\left(\frac{g_{N\omega}^2}{m_\omega^2}[f(x)+xf'(x)]+\frac{g_{N\phi}^2}{m_\phi^2}\right) \, .
\end{equation}

The values of the hyperon potential depths which are given, explicitly
\begin{equation}\label{potentialdepths}
    U_{i}= m_{i,+}(\sigma_{0})-  m_{i,+}(f_{\pi}) +g_{i \omega} \omega_{0}+ g_{i \phi} \phi_{0} \, , \qquad (i=\Lambda,\Sigma,\Xi).
\end{equation}

We can now solve the coupled system of equations that contains: Eqs. (\ref{Psat}), (\ref{sigsat}), (\ref{gNw}), (\ref{incomp}), (\ref{gNrho}), (\ref{L}), the three Eqs. (\ref{potentialdepths}) and the mass constraints of $3$ vacuum masses $m_{N,+}^{\rm vac} = 939\, {\rm MeV}$, $m_{\Sigma,+}^{\rm vac} = 1190\, {\rm MeV}$ and $m_{\Lambda,+}^{\rm vac} = 1116\, {\rm MeV}$, in terms of the parameters: $a_2$, $a_3$, $a_4$, $d$, $g_{N\omega}$, $g_{N\rho}$, $g_{N \phi}$, $g_{N \sigma}^{(1)}$, $g_{\Sigma \sigma}^{(1)}$, $g_{\Lambda \sigma}^{(1)}$, $m_{1}$, $m_{2}$.

\bibliography{references}

\end{document}